\documentclass[letterpaper,useAMS,usenatbib]{mn2e}

\usepackage{graphicx}
\usepackage{graphics}
\usepackage{epstopdf}
\usepackage{amsmath}
\usepackage{amssymb}
\usepackage{bm}
\usepackage{verbatim}

\newcommand{\Hb}{H$\beta$}

\begin{document}

\title[Modeling reverberation mapping data I]{Modeling reverberation mapping data I: \\
improved geometric and dynamical models and comparison with cross-correlation results}

\author[A. Pancoast et al.]
{Anna Pancoast$^1$,
Brendon J. Brewer$^2$,
Tommaso Treu$^{1, 3}$ \\
$^1$Department of Physics, University of California, Santa Barbara, CA 93106, USA; pancoast@physics.ucsb.edu \\
$^2$Department of Statistics, The University of Auckland, Private Bag 92019, Auckland 1142, New Zealand \\
$^3$Current address: Physics and Astronomy Building, 430 Portola Plaza, Box 951547, Los Angeles, CA 90095-1547, USA
}

\maketitle

\begin{abstract}
We present an improved and expanded simply parameterized
phenomenological model of the broad line region (BLR) in active
galactic nuclei (AGN) for modeling reverberation mapping data.  By
modeling reverberation mapping data directly, we can constrain the
geometry and dynamics of the BLR and measure the black hole mass without
relying on the normalization factor needed in the traditional analysis.  
For realistic simulated reverberation
mapping datasets of high-quality, we can recover the black hole mass
to $0.05-0.25$ dex uncertainty and distinguish between dynamics dominated by
elliptical orbits and inflowing gas.  While direct
modeling of the integrated emission line light curve allows for
measurement of the mean time lag, other details of the geometry of the
BLR are better constrained by the full spectroscopic dataset of
emission line profiles.  We use this improved model of the BLR to
explore possible sources of uncertainty in measurements of the time
lag using cross-correlation function (CCF) analysis and in
measurements of the black hole mass using the virial product.    
Sampling the range of geometries and dynamics
in our model of the BLR suggests that the theoretical uncertainty in black hole
masses measured using the virial product is on the order of 0.25 dex.
These results support the use of the CCF to measure
time lags and the virial product to measure black hole masses when
direct modeling techniques cannot be applied, provided the uncertainties 
associated with the interpretation of the results are taken into account.
\end{abstract}


\section{Introduction \label{sect_intro}}

While active galactic nuclei (AGN) are thought to be powered by
accretion onto super-massive black holes \citep{lynden71}, much
remains unknown about the physical distribution and kinematics of the
surrounding gas.  
A portion of this gas is deep enough in the potential of the black
hole that its emission lines are broadened by up to several thousand
km/s \citep{antonucci93, urry95}.  At distances from the black hole of
$\sim 10^{14}-10^{16}$ m \citep{wandel99, kaspi00, bentz06, bentz13},
this so-called broad line region provides a unique probe of AGN
physics and the opportunity to measure the mass of the black hole
itself, via reverberation mapping \citep{blandford82, peterson93,
peterson04}.

Traditionally, the goal of reverberation mapping is to infer a
characteristic radius of the BLR by measuring the time lag between
changes in the AGN ionizing continuum (or a proxy such as the optical
AGN continuum) and the response of the broad emission line flux.  If
the time lag is due only to light travel time, then the characteristic
radius of the BLR is given by the speed of light $c$ times the time
lag $\tau$.  The time lag can then be combined with a measure of the
velocity of the BLR gas to form a dimensional black hole mass estimate
referred to as the virial product, $M_{\rm vir} = c\tau \Delta v^2/G$,
where $G$ is the gravitational constant and $\Delta v$ is a measure of
the width of the broad emission line profile.  The virial product is related to the
true black hole mass $M_{\rm BH}$ by the virial coefficient $f$, a
dimensionless factor of the order unity.  In recent years, the
standard practice is to set the average virial coefficient by
aligning the $M_{\rm BH} -
\sigma_*$ relations for galaxies with dynamical $M_{\rm BH}$
measurements and galaxies with reverberation mapped black hole mass
measurements \citep{onken04, collin06, woo10, greene10b, graham11,
park12b, woo13, grier13b}.  The scatter in the $M_{\rm BH} - \sigma_*$
relation of $\sim 0.4$ dex
\citep[e.g.][]{park12b} indicates that the uncertainty introduced by
assuming a single value of $f$ for the full reverberation mapped
sample could be comparable to the intrinsic scatter of the $M_{\rm BH}
- \sigma_*$ for quiescent galaxies.

With a new generation of high-quality reverberation mapping datasets
\citep{bentz09, denney10, barth11, grier13a}, more information 
has become available to probe the details of the geometry and dynamics
of the BLR. In order to exploit this improvement in the data, in the
last few years we have been developing a new technique for
analyzing reverberation mapping data \citep{pancoast11}. The
fundamental difference of our approach with respect to previous work
is that we aim to fit fully self-consistent models of the BLR
geometry and dynamics to the data, rather than trying to infer
the so-called transfer function \citep{blandford82, horne91, horne94, krolik95}.  
In addition to providing quantitative
constraints on the geometry and dynamics of the BLR, this direct
modeling approach allows us to measure the black hole mass without
relying on the virial coefficient $f$; instead, the black hole mass is just one
of the parameters in our model fit.  Previously we applied this
direct modeling approach to the \Hb\ emission line in two AGNs, Arp
151 \citep{brewer11} and Mrk 50
\citep{pancoast12}, that were observed as part of the Lick AGN
Monitoring Projects in 2008 \citep{walsh09, bentz09} and 2011
\citep{barth11}, respectively.

In this paper, the first of a series on direct modeling of reverberation
mapping data, we introduce an improved and expanded version of our
simply parameterized phenomenological model of the BLR to be used with
the direct modeling approach.  We demonstrate the capabilities of this
new model using simulated data and by placing constraints on the
uncertainties in traditional cross-correlation function (CCF)
analysis.  In paper II of this series \citep{pancoast14}, we apply the
improved BLR model to five AGNs in the LAMP 2008 dataset.  The
additional model flexibility and increased algorithm efficiency of
this new implementation are demonstrated by comparing the results for
Arp 151 by \citet{brewer11} to the new results described in paper II;
in the latter case the uncertainty in black hole mass is decreased by
more than 0.1 dex and it is possible to differentiate between inflow
and outflow kinematics.

We begin by presenting a detailed description of the improved BLR
model in Section~\ref{sect_model}.  Tests to recover the model
parameters using simulated data are presented in
Section~\ref{sect_simdata_tests}.  Comparison of direct modeling
results to CCF analysis and constraints on CCF lag uncertainties are
given in Section~\ref{sect_ccf_tests}.  Finally, we give an overview
of the main conclusions in Section~\ref{sect_conclusions}.  Throughout
this paper, all BLR model parameter values are given in the rest frame
of the AGN.

\section{The Model}
\label{sect_model}

In this section we describe our model of the BLR and the numerical
methods we use to explore its parameter space.  Our model of the
BLR can be applied to any broad emission line, although it has so far
only been applied to the \Hb\ broad emission line in six AGNs
\citep{brewer11, pancoast12, pancoast14}.  The basic methodology of
our model is also completely generalizable to any model in which the
geometry and dynamics of the BLR gas can be computed quickly enough to
enable a full exploration of the parameter space when comparing with
the data.

\subsection{Overview}
Our goal is to reconstruct the physical structure of the BLR and to measure the mass
of the central black hole from reverberation
mapping measurements. To achieve this, we describe the possible structure of
the BLR by a large number of parameters whose values we infer from the data.

In our model, the BLR is represented by a set of point particles whose positions
represent the spatial distribution of broad line emission.
If the BLR is really made up of distinct clouds, then each particle could be
associated with emission from a BLR cloud, however if the BLR is made up of a
smoother distribution of gas, then the particles are just a
Monte Carlo approximation of the density field of emission. Each particle in
our model is also associated with a velocity that depends upon
the mass of the black hole. Our model parameters for the BLR describe the
spatial distribution of the particles as well as their individual positions.
Additional parameters describe the rule by which velocities are assigned to
the particles, as well as the individual velocities themselves.
In the present implementation we ignore
gravitational interactions or fluid viscosity between particles,
and other non-gravitational forces like radiation pressure.

Given a distribution of particles with associated velocities, we can
immediately calculate how the BLR would process an input continuum light
curve, resulting in an emitted broad line spectrum (e.g. \Hb) that changes
(in both total flux and shape) over time.  Apart from the conversion from
continuum to line flux, we assume that the particles act as
mirrors, reflecting the continuum flux towards the observer, where the
velocity of the particle determines how far the emission line
flux is shifted in wavelength space away from the systematic emission
line wavelength at rest with respect to the black hole.

There are three parts to our model of the BLR, which is formulated as
an application of Bayesian inference as described in
Section~\ref{sect_bayes}.  The first part of the model is the AGN
continuum light curve model described in Section~\ref{sect_continuum}.
It is necessary to model the AGN continuum light curve because we need
to be able to evaluate the continuum light curve at arbitrary times in
order to calculate the broad line spectrum variations predicted by the model.
The second part is the ``geometry model'' (spatial distribution) of the BLR
described in Section~\ref{sect_geomodel}, which describes the spatial
distribution of the particles that make up the BLR emission.
The positions of the particles
determine their time lags, which tells us how delayed features in the broad emission
line light curve are compared to the continuum light curve.  The third
part is the ``dynamical model'' of the BLR described in
Section~\ref{sect_dyn}. This describes the rule by which velocities are
assigned to the particles, and allows for scenarios such as near-circular
orbits, inflow, or outflow.
The component of a particle's velocity along the line of sight determines
which wavelength it affects in the model-predicted broad line spectrum.
Once the three parts of our
model of the BLR have been specified, we must explore the model
parameter space in order to constrain the properties of the BLR given a
specific reverberation mapping dataset, as described in
Section~\ref{sect_dnest}.  Finally, we enumerate the limitations of
our current model of the BLR and future improvements in
Section~\ref{sect_lim}.

\subsection{Bayesian Inference Framework}
\label{sect_bayes}
We use the formalism of Bayesian statistics to infer the values of our
model parameters $\boldsymbol{\theta}$ given a reverberation mapping
dataset ${\bf D}$.  We begin by defining the prior probability distributions of
the model parameters, $p(\boldsymbol{\theta} | I)$, which incorporate our
initial assumptions about the range of allowed parameter values and
depend upon any information $I$ that we have about the problem before
we begin.  We then assign the probability distribution of the data given
a specific set of parameter values $p({\bf D} | \boldsymbol{\theta}, I)$
which tells us how the data and model parameters are related. This term is 
often called the ``sampling distribution'', or, once the data is known, the
likelihood.
Finally, we can combine the prior and likelihood using Bayes'
theorem to obtain the posterior distribution of the
model parameters given the data:
\begin{equation}
\label{eqn_bayes}
p(\boldsymbol{\theta} | {\bf D}, I) \propto p(\boldsymbol{\theta} | I)\, p({\bf D} | \boldsymbol{\theta}, I).
\end{equation}

The normalization constant of the posterior in Equation~\ref{eqn_bayes},
called the {\it evidence} or the {\it marginal likelihood}, is given by
\begin{eqnarray}
p({\bf D} | I) &=& \int p(\boldsymbol{\theta} | I)\, p({\bf D} | \boldsymbol{\theta}, I) \, d^n{\bf\theta}
\end{eqnarray}
and is useful for model comparison.

For models with many parameters and in which the posterior distribution
is not of a known standard form, it is common to calculate properties of the
posterior
probability density function (PDF) by generating samples
using an algorithm such as Markov Chain Monte
Carlo (MCMC).  As the number of parameters becomes large and the
likelihood function potentially multimodal, however, it can be more efficient to
use a more complex algorithm such as Diffusive Nested Sampling (DNS), as
described in Section~\ref{sect_dnest}.  DNS has
the added benefit that it computes the marginal likelihood,
allowing for model selection, unlike most standard MCMC algorithms that only
generate posterior samples.

In our inference problem of modeling the BLR, the data consist of two
time series. The first is the AGN continuum light curve $\{\mathcal{Y}_i\}$
and its corresponding timestamps and measurement error
variances. The second time series is the spectrum of the broad line
measured over time, which we will denote by $\{\mathcal{D}_{ij}\}$ (the index
$i$ represents the epoch and $j$ the wavelength bin). The overall
dataset that enters into Bayes' theorem is both of these:
\begin{equation}
{\bf D} = \{ \{\mathcal{Y}_i\}, \{\mathcal{D}_{ij}\} \}.
\end{equation}
We can split the likelihood function into two parts. The likelihood for the
continuum data $\{\mathcal{Y}_i\}$ will be discussed in
Section~\ref{sect_continuum}. For the broad line data,
we use the model parameters $\boldsymbol{\theta}$ to construct a time series
of mock broad emission line spectra $m_{ij}(\boldsymbol{\theta})$ to compare to
the data using a Gaussian likelihood function:
\begin{equation}
p({\bf D} | \boldsymbol{\theta}, I) =  \prod_{i,j} \frac{1}{\sigma_{ij}  \sqrt{2 \pi}}
 \exp \left[   -\frac{1}{2\sigma_{ij}^2} \left(\mathcal{D}_{ij} - m_{ij}(\boldsymbol{\theta})\right)^2
 \right]  
\end{equation}

\subsection{Continuum Light Curve Model}
\label{sect_continuum}
Ground-based reverberation mapping campaigns use optical AGN continuum
light curves (e.g. in the {\it V} or {\it B} bands) to track the
variability of photons leading to BLR emission, since the true
ionizing photons are in the ultraviolet (UV).  While it is expected
that the UV photons are created in the accretion disk closer to the
black hole than the optical photons, the time lag between variability
features in the UV and optical is unresolved \citep{peterson91,
korista95} or on the order of a day \citep{collier98}.  For this
reason, we do not distinguish between a UV or optical light curve in
our model of the BLR, assuming that either light curve is emitted from
a point source at the position of the black hole.  While the true UV
and optical emitting regions in the accretion disk are certainly not
point-like, their distance from the black hole is significantly
smaller than that of the BLR compared to the uncertainties
in the mean BLR radius \citep[e.g.][]{morgan10}, suggesting that
detailed modeling of the optical or UV emitting region would not be
well-constrained by current reverberation mapping datasets.
Since our model of the BLR is many particles each reflecting the
continuum light curve to the observer with a time lag given by the
particle's distance from the continuum point source, the
continuum flux must be computed at arbitrary times within the light
curve.  Generally, reverberation mapping AGN continuum light curves
are too sparsely sampled to resolve intra-day variability using simple
linear interpolation between data points.  Linear interpolation also
incorrectly assumes that there is no uncertainty associated with the
interpolation process or the measurements. 
For these reasons, we model the AGN continuum
light curve using a stochastic model of AGN variability, allowing us
to evaluate the light curve at arbitrarily small timescales and also to
include the continuum light curve model uncertainty into our inference
on the properties of the BLR.

We model the continuous AGN continuum light curve $y(t)$ using Gaussian
processes (GPs), which
allow us to treat the interpolated and extrapolated light curve points as additional
parameters in our model, constrained by the data $\bf D$. Most of the
information about $y(t)$ is, as one would expect, provided by the continuum
light curve data $\{\mathcal{Y}_i\}$.

With the GP assumption, the prior distribution for any finite set of
interpolated flux values is a multivariate Gaussian:
\begin{eqnarray}
p(\mathbf{y} | \mu_{\rm cont}, \mathbf{C}) &=& \frac{1}{\sqrt{(2 \pi)^n \det \mathbf{C}}}
\times\\
& &
\exp \left[  
-\frac{1}{2}  (\mathbf{y} - \mu_{\rm cont})^T {\bf C}^{-1} (\mathbf{y} - \mu_{\rm cont})
\right]
\end{eqnarray}
where $\mathbf{y}$ are the interpolated continuum light curve
points (i.e. evaluations of the function $y(t)$),
${\mu_{\rm cont}}$ is the long-term mean flux value of the
light curve, and ${\bf C}$ is the covariance matrix.  The covariance
between any two points in the interpolated continuum light curve
depends on the time difference between them, as given by:
\begin{equation}
C(t_1, t_2) = \sigma_{\rm cont}^2 \exp \left[  - \left(  \frac{| t_2 - t_1 |}{\tau_{\rm cont}} \right)^{\alpha_{\rm cont}} \right]
\end{equation}
where $\sigma_{\rm cont}$ is the long term standard deviation of the
continuum light curve, $\tau_{\rm cont}$ is the typical timescale for
variations, and $\alpha_{\rm cont}$ is a smoothness parameter between
1 and 2.  Larger values of $\alpha_{\rm cont}$ lead to more covariance
between points in the continuum light curve, corresponding to less
fluctuations on small timescales.  Setting $\alpha_{\rm cont} = 1$
improves the speed with which the densely sampled continuum light
curve can be calculated, as well as increasing the performance
of the MCMC\footnote{Performance is also increased by parameterising in terms
of $\sigma_{\rm cont}/\sqrt{\tau_{\rm cont}}$ rather than $\sigma_{\rm cont}$ itself.}.
For these reasons, we generally
set $\alpha_{\rm cont} = 1$, in which case our Gaussian process model
is equivalent to a continuous time first-order autoregressive process
(CAR(1)).  The CAR(1) model has been found to be a good fit to AGN
variability data on similar timescales to those probed by reverberation mapping campaigns
\citep{kelly09, kozlowski10, macleod10, zu11, zu13}, although a model
that further reduces AGN variability on very short timescales provides a better fit to higher-cadence Kepler data \citep{mushotzky11}. 
We interpolate and extrapolate the AGN continuum light
curve data using 1000 points, where the range of points starts before
the continuum data (usually by 50\% the continuum data range) and
extends past the end of both the continuum and line data, whichever is
later.  Points extrapolated past the ends of the continuum data are
only constrained by the general behavior of the interpolated points
and thus have very high uncertainty.

\subsection{Geometry Model}
\label{sect_geomodel}
Once we have a model for the continuum light curve we need a model for
the spatial distribution of the particles, which we call the ``geometry
model''.  The geometry model has flexibility in the radial distribution
of the particles as well as the angular distribution. In particular we include an
opening angle parameter that describes whether the BLR is a disk or
sphere and an inclination angle parameter that determines from what
angle the observer sees any asymmetries of the BLR. Although this is a
purely phenomenological model, it is flexible enough that it should
allow us to capture a wide variety of realistic geometries with a
moderate number of parameters.
\begin{figure}
\begin{center}
\includegraphics[scale=0.67]{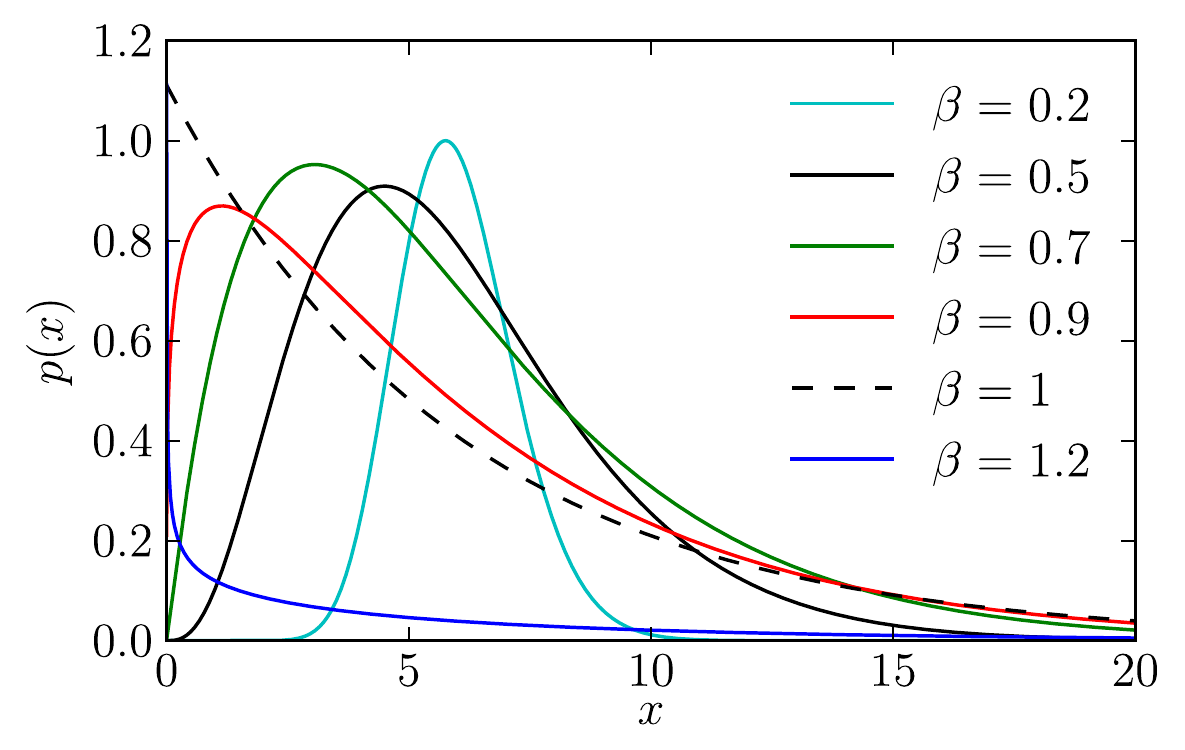}
\caption{ Examples of possible radial profiles for the BLR emission
given by the Gamma distribution with $\mu = 6$, $F = 0$, and
various values for $\beta$.  The distributions range from a narrow
Gaussian ($\beta < 1$) to an exponential profile ($\beta = 1$) or
steeper ($\beta > 1$).}
\label{fig_radialprof}
\end{center}
\end{figure}

We define the geometry model in two stages. First we consider the radial
distribution of the particles, and secondly we define the angular structure.
\subsubsection{Radial BLR Distribution}
The radial distribution of BLR emission density is described by a
shifted gamma distribution. The gamma distribution for a positive
variable $x$ is usually written
\begin{equation}
p(x|\alpha,\theta) \propto x^{\alpha-1} \exp \left( -\frac{x}{\theta} \right)
\end{equation}
where $\alpha$ is the shape parameter and $\theta$ is a scale parameter.
Our radial distribution is based on a shifted gamma distribution where the
lower limit is $r_0$ instead of zero. Rather than parameterizing the
distribution by $(\alpha, \theta, r_0)$, whose interpretations are not
straightforward (making priors difficult to assign),
we use a different parameterisation in terms of
three parameters $(\mu, \beta, F)$, defined as follows.

\begin{eqnarray}
 \mu &=& r_0 + \alpha\theta \label{eqn_mu} \\
 \beta &=& \frac{1}{\sqrt{\alpha}} \label{eqn_beta}\\
 F &=& \frac{r_0}{r_0+\alpha\theta}. \label{eqn_f}
\end{eqnarray}

The parameter
$\mu$ is the mean value
of the shifted gamma distribution, $\beta$ is the standard
deviation of the gamma distribution in units of the mean $\mu$ when
$r_0=0$, and $F$ is the fraction of $\mu$ from the origin at which
the gamma distribution begins (i.e. $F$ is $r_0$ measured in units of $\mu$).  
 For arbitrary $r_0$, the standard deviation of the shifted gamma distribution is:
\begin{eqnarray}
\sigma_r &=& \mu \beta (1 - F).
\end{eqnarray}
Finally, we also offset the radial distribution by the Schwarzschild
radius, $R_s = 2GM/c^2$, to provide a hard limit to how close a point
particle can be to the black hole.  For a $10^7 M_\odot$ black hole,
$R_s = 0.001$ light days, much smaller than the typical size of the
BLR, which is on the order of light days.

The three parameters $(\mu, \beta, F)$ control the radial profile of the
particles. To parameterise the actual distances of the particles
from the black hole, instead of using the physical distance $r$, we use a variable
$g$ (one for each particle) with a Gamma$(\beta^{-2}, 1)$ prior.
Then the actual distance $r$ of the particle is computed by:
\begin{eqnarray}
r &=& R_s + \mu F + \mu\beta^2(1-F)g.\label{eqn_calculate_r}
\end{eqnarray}
The reason for parameterizing in terms of $g$ rather than $F$ is that
Metropolis proposals are simpler. For example, a Metropolis move that changes
the parameters $(\mu, \beta, F)$ but leaves $g$ fixed
will automatically move all of the particles
appropriately.

\subsubsection{Opening and Inclination Angles}
\label{sect_angles}
The radial BLR distribution discussed in the previous section is
spherically symmetric, however we can break spherical symmetry by
introducing a disk opening angle of the BLR. The opening angle is
defined as half the angular thickness of the BLR in the angular
spherical polar coordinate perpendicular to the plane of the disk. If
the BLR is a sphere then the opening angle is $\pi/2$, and if the BLR
is a thin disk then the opening angle approaches zero. Once spherical
symmetry has been broken, it is necessary to consider at what
angle an observer will view the BLR.  The inclination angle is defined
as the angle between a face-on BLR geometry and the observer's line of
sight, so an edge-on disk would have an inclination angle of $\pi/2$
while a face-on disk would have an inclination angle approaching zero.

 To construct a specific BLR geometry, we begin by drawing the radial
position for each particle in a flat disk in the $x$-$y$ plane
with the observer located at the positive end of the $x$-axis.
In plane polar coordinates, the radial coordinates $r$ of the point
particles are calculated using Equation~\ref{eqn_calculate_r}, and the angular
coordinates are drawn from a uniform distribution between 0 and $2\pi$.
We then puff up this flat disk
by the opening angle, first by rotating each particle around the
$y$-axis by some angle between 0 and the opening angle and then by
rotating the particle around the $z$-axis by some angle between
0 and $2\pi$ to restore axisymmetry. Next, we rotate all point
particles around the $y$-axis by 90 degrees minus the inclination
angle so that an inclination angle of zero corresponds to a face-on
BLR geometry. All of the angles used in this process are extra model
parameters.

\subsubsection{Angular BLR Distribution}
\label{sect_angular}
We can further add asymmetry by controlling the strength of emission
from a given particle using three separate effects:
\begin{enumerate}
 \item  The particles are assigned non-uniform weights, depending upon 
 the angle between the observer's line of sight to the central source and a particle's
 line of sight to the central source. The strength of this effect is controlled
 by a parameter $\kappa$.
 \item The parameter $\gamma$ controls the extent to which the emission is 
 concentrated near the outer edges of the BLR disk at the opening angle.
 \item The parameter $\xi$ determines the transparency of the plane of the BLR disk.
\end{enumerate}

The first effect represents anisotropic emission from the point
particles.  We use first order spherical harmonics to define a weight,
$W$, for each particle that ranges from 0 to 1 and determines
what fraction of the continuum flux is reemitted as line flux in the
direction of the observer:
\begin{equation}
 W(\phi) = \frac{1}{2} + \kappa \cos \phi. \label{eqn_kappa}
\end{equation}
The one free parameter is $\kappa$, which ranges from $-0.5$ to $0.5$.
Negative values of $\kappa$ correspond to preferential emission from
the far side of the BLR from the observer and positive values
correspond to preferential emission from the near side of the BLR.
Preferential emission from the far side of the BLR could be physically
caused by BLR gas only re-emitting continuum emission back towards the
central source due to self-shielding, while preferential emission from
the near side of the BLR could be physically caused by the closer BLR
gas blocking gas farther away.  The angle $\phi$ is defined to be the
angle between the observer's line of sight to the central source and
the particle's line of sight to the central source.  For $\kappa
= -0.5$ and a model where the BLR is made up of spherical balls of
gas, this model is equivalent to considering broad line emission from
the area of the spheres illuminated by the central source as viewed by
the observer, like lunar phases.

The second effect is parameterized by $\gamma$ and controls the extent
to which BLR emission is concentrated near the outer faces of a disk.
This could arise for example if the parts of the BLR closer to the
plane of the accretion disk are optically thick.
The parameter $\gamma$ controls preferential emission from the outer
faces of the BLR disk by affecting how much the particle
positions are moved from an initial flat disk to between zero and the
opening angle of a thick disk.  The angle for a particle's
displacement from a flat to thick disk is given by:
\begin{equation}
 \theta = {\rm acos} \left( \cos \theta_o + (1 - \cos \theta_o)\times U^\gamma \right)  \label{eqn_gamma}
\end{equation}
where $\theta_o$ is the opening angle and $U$ is a random number drawn
uniformly between 0 and 1. Larger values of $U$ lead to $\theta$
values closer to $\theta_o$, so using $U^\gamma$ with $\gamma$ between
1 and 5 concentrates more particles close to the opening angle
for $\gamma > 1$.

The third effect represents the possibility for an obscuring medium in
the plane of the BLR to partly or completely obscure broad line
emission from the back side of the BLR and is parameterized by $\xi$.
Unlike the first effect that depends upon the inclination angle at
which an observer views the BLR, $\xi$ is roughly defined as the
fraction of particles on the far side of the BLR midplane.  In
the limit of $\xi \to 0$, the entire back half of the BLR is obscured,
and the BLR geometry could range from half a disk or sphere when
$\gamma \sim 1$ to a single cone when $\gamma \sim 5$.  In the limit
of $\xi \to 1$, the back half of the BLR is not obscured.  Since it is
computationally inefficient to throw out particles on the back
side of the BLR, we actually just invert their position with respect
to the plane of BLR when $\xi < 1$, making the true definition of
$\xi$ be the fraction of particles in the back side of the BLR
that have {\it not} been moved to the front side.

\subsection{Dynamics Models}
\label{sect_dyn}
In order to make a model spectrum from our geometry of the BLR we must
also assign velocities to the particles.  We consider three
different kinematic components, including bound elliptical orbits and
a combination of both bound and unbound inflow or outflow.

\begin{figure}
\begin{center}
\includegraphics[scale=0.44]{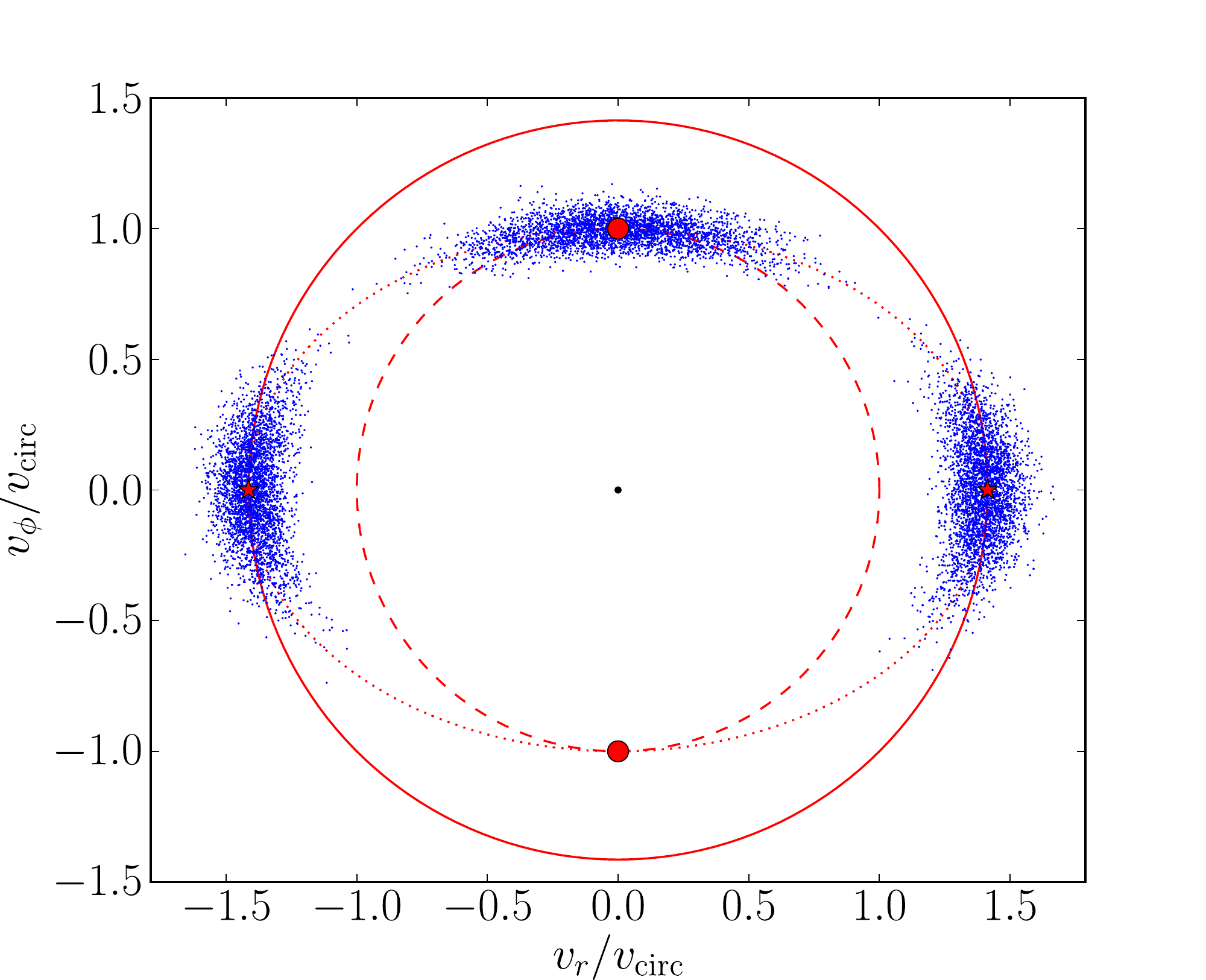}
\caption{Distributions of radial and tangential velocities, $v_r$ and $v_\phi$ for the dynamical model.  
Blue points are particle velocities drawn from Gaussian
distributions centered around the point for circular orbits $(v_r,
v_\phi) = (0, v_{\rm circ})$ shown as the upper filled red circle and
centered around the points for outflowing and inflowing escape
velocity $(v_r, v_\phi) = (\pm \sqrt{2}\,v_{\rm circ}, 0)$ shown as
filled red stars.  The red dotted line denotes the ellipse with
semi-minor axis $(v_r, v_\phi) = (0, \,v_{\rm circ})$ and semi-major
axis $(v_r, v_\phi) = (
\sqrt{2}\,v_{\rm circ}, 0)$ along which the radial and tangential
velocities are drawn.  The outer solid red circle at a radius of $|v|
= \sqrt{2GM_{\rm BH}/r}$ denotes the velocity beyond which orbits are unbound.
The red dashed circle at a radius of $|v| = \sqrt{GM_{\rm BH}/r}$ denotes
velocities with magnitude of the circular velocity.}
\label{fig_newVmodel}
\end{center}
\end{figure}

\subsubsection{Elliptical Orbits}
\label{sect_ellip}
Consider a particle orbiting a point mass at a distance $r$ with
velocity $|v| = \sqrt{v_r^2 + v_\phi^2}$, where $v_r$ is the radial
velocity and $v_\phi$ is the tangential velocity in the plane of the
orbit and perpendicular to $v_r$.  The tangential velocity in terms of
the angular momentum per unit mass of the particle $L$ is given by
$v_\phi = L/r$, and the radial velocity can be obtained by considering
the energy per unit mass of the particle:
\begin{equation}
 E = \frac{1}{2} \left( v_r^2 + \frac{L^2}{r^2} \right) - \frac{GM_{\rm BH}}{r}. \label{eqn_m}
\end{equation}
Solving for $v_r$ we obtain:
\begin{equation}
 v_r = \pm \sqrt{2 \left( E + \frac{GM_{\rm BH}}{r} \right) - \frac{L^2}{r^2}}.
\end{equation}
For circular orbits, we have the additional constraint that $v_r=0$ so
that the centripetal force of circular motion must equal the
gravitational force, giving $v_\phi^2 = GM_{\rm BH}/r$ or $v_{\rm circ} =
\sqrt{GM_{\rm BH}/r}$.  Thus, the circular orbit solutions are two special
points in the $v_r - v_\phi$ plane at $(v_r, v_\phi) = (0, \pm v_{\rm
circ})$.

We consider generalizations of circular orbits to elliptical orbits by
considering distributions in $v_r$ and $v_\phi$ centered around the
circular orbit solutions.  Such a model allows us to recover circular
orbits when the distributions are narrow, but also allows for highly
elliptical orbits when the distributions are on the order of $v_{\rm
circ}$.  We draw the velocities of the particles from the
ellipse in the $v_r$ and $v_\phi$ plane that has semi-minor axis
$v_{\rm circ}$ at $v_r=0$ and semi-major axis equal to the escape
velocity $\sqrt{2}v_{\rm circ}$ at $v_\phi=0$, as shown in
Figure~\ref{fig_newVmodel}.  The reason for drawing velocities from
around this ellipse instead of a circle with radius $v_{\rm circ}$ is
that the parameter space naturally includes the points at $v_r = \pm
\sqrt{2}v_{\rm circ}$ that correspond to the radial outflowing and
inflowing escape velocities.  We will discuss these inflowing and
outflowing velocity solutions in more detail in
Section~\ref{sect_inflow}.  Since reverberation mapping measurements
cannot distinguish between rotations of the BLR around the line of
sight axis, it is only necessary to define the positive $v_\phi$ side
of the $v_r - v_\phi$ plane.  The radial and tangential velocities are
thus drawn from Gaussian distributions centered at $(v_r, v_\phi) =
(0, v_{\rm circ})$ with standard deviations given by
$\sigma_{\rho,\,\rm circ}$ and $\sigma_{\Theta,\,\rm circ}$, where
$\rho$ is the radial coordinate in the $v_r - v_\phi$ plane and
$\Theta$ is the angular coordinate.  Circular orbits are recovered
when $\sigma_{\rho,\,\rm circ} \to 0$ and $\sigma_{\Theta,\,\rm circ}
\to 0$, whereas highly elliptical orbits approaching the escape velocity
$|v| = \sqrt{2}v_{\rm circ}$ are obtained when $\sigma_{\rho,\,\rm
circ} \to 0.1$ and $\sigma_{\Theta,\,\rm circ} \to 1.0$, the upper
limits of their priors.

\subsubsection{Inflow and Outflow}
\label{sect_inflow}
In order to include the possibility of substantial unbound outflowing
or inflowing gas in the BLR, we allow a variable fraction of the point
particles to have elliptical, inflowing, and outflowing orbits.  Since
we do not expect to find both inflowing and outflowing gas in the BLR
in the same spatial location, especially at the velocities assumed by
our model, we only allow for inflowing or outflowing particles
in addition to elliptical orbits for a specific instance of our model.
The fraction of particles with elliptical orbits is given by
$f_{\rm ellip}$, where $1 - f_{\rm ellip}$ is thus the fraction of
particles in either inflowing or outflowing orbits. Whether the
orbits are inflowing or outflowing is given by $f_{\rm flow}$, where
values between 0 and 1 and less than 0.5 denote inflow and values
greater than 0.5 denote outflow.  Inflowing orbits are obtained around
values of $(v_r, v_\phi) = (-\sqrt{2}\,v_{\rm circ}, 0)$ while
outflowing orbits are obtained around values of $(v_r, v_\phi) =
(\sqrt{2}\,v_{\rm circ}, 0)$.

As for elliptical orbits, we draw the radial and tangential velocities
of inflowing or outflowing particles from Gaussian distributions
for $\rho$ and $\Theta$, the radial and angular coordinates of the $v_r
- v_\phi$ plane.  The width of the Gaussian distributions is similarly
given by $\sigma_{\rho,\,\rm radial}$ and $\sigma_{\Theta,\,\rm
radial}$, where the widths are the same for both inflowing and
outflowing orbits.  Since the Gaussian distributions are centered on
the points $v_r = \pm \sqrt{2}\,v_{\rm circ}$, about half of the
inflowing and outflowing particles will actually have bound
orbits.  In order to allow for completely bound inflowing and
outflowing trajectories, we also allow the distributions centered
around $v_r = \pm \sqrt{2}\,v_{\rm circ}$ to be rotated by an angle
$\theta_e$ along the ellipse connecting $v_r = \pm \sqrt{2}\,v_{\rm
circ}$ to the circular orbit velocities $v_\phi = \pm v_{\rm circ}$.
When $\theta_e = 0$, the inflowing or outflowing orbits are centered
around the escape velocities at $v_r = \pm \sqrt{2}\,v_{\rm circ}$,
while $\theta_e \to \pi/2$ recovers bound elliptical orbits centered
around circular orbits.  When $\theta_e \sim \pi/4$, we obtain mostly
bound inflowing or outflowing gas.

\subsubsection{Macroturbulent Velocities}
\label{sect_turbulence} 
We also consider macroturbulent velocities of the particles in
addition to the velocities from elliptical, inflowing, or outflowing
orbits.  For each particle, we calculate the magnitude of the
turbulent velocity along the observer's line of sight, given by:
\begin{equation}
 v_{\rm turb} = \mathcal{N}(0,\sigma_{\rm turb}) |v_{\rm circ}|
\end{equation}
where $\mathcal{N}(0,\sigma_{\rm turb})$ is a normal distribution
centered on zero and with standard deviation $\sigma_{\rm turb}$. The
magnitude of the turbulent velocity is relative to the magnitude of
the velocity of the particle's circular orbit described in
Section~\ref{sect_ellip}, given by $v_{\rm circ}$.  We can recover the
case with no additional turbulent velocities when $\sigma_{\rm turb}
\to 0$.  We apply the additional macroturbulent velocity to a point
particle first by calculating the elliptical, inflowing, or outflowing
velocity and then adding $v_{\rm turb}$.  This model for
macroturbulent velocities is similar to the one presented by
\citet{goad12} for the case of a disk with constant opening angle.

\subsubsection{Relativistic Effects}
As highlighted in \citet{goad12}, relativistic effects can have a
strong influence on the shape of emission line profiles if the BLR gas is
sufficiently close to the black hole.  We include two simple
relativistic effects in the calculation of particle velocities.
The first effect is the full relativistic expression for the doppler
shift of the broad emission line due to the line of sight velocity of
the emitting BLR gas.  The second relativistic effect is that of
gravitational redshift, which is caused by a photon being emitted from
deeper in a gravitational potential well than the observer of the
photon.  The wavelength shift caused by gravitational redshift depends
upon the ratio of the Schwarzschild radius, $R_s = 2GM/c^2$, to the
radial distance of the emitting source.  Together, the full
relativistic expression for doppler shift and the expression for
gravitational redshift act to shift the emitted wavelength
$\lambda_{\rm emit}$ of line emission from a particle to the
observed wavelength $\lambda_{\rm obs}$ given by:
\begin{equation}
 \label{eqn_gravred}
 \lambda_{\rm obs} = \lambda_{\rm emit} \frac{\sqrt{\frac{1 + \frac{v}{c}}{1 - \frac{v}{c}}}}{\sqrt{1 - \frac{R_s}{r}}} 
\end{equation}
where the particle has velocity $v$ and radial distance from the
black hole $r$.  Since we compare our model broad emission line
spectra to the data in wavelength space, we can include the
relativistic doppler shift and gravitational redshift in the simulated
data by converting the simulated data from velocity to wavelength
space using Equation~\ref{eqn_gravred}.

\subsubsection{Narrow Line Emission}
In addition to a model of the broad emission line, we must also
consider the superimposed narrow emission line from the narrow line
region (NLR).  Since the NLR is farther from the black hole, the
narrow emission line is not expected to reverberate on timescales as
short as those for the BLR \citep[e.g.][]{peterson13}.  We therefore
assume that the narrow emission line flux is constant over the
duration of a reverberation mapping dataset.  
We model the narrow emission line component using a Gaussian 
with line dispersion given by another more
isolated narrow emission line profile.
For example, to model the narrow
component of the \Hb\ emission line we use the line dispersion of the narrow [\mbox{O\,{\sc iii}}]$\lambda5007$ 
emission line, just red-ward of \Hb.  
Since the width of [\mbox{O\,{\sc iii}}]$\lambda5007$
in a given reverberation mapping dataset
is due to both intrinsic line width and instrumental resolution, we use
measurements of the intrinsic line width to calculate the instrumental 
resolution, which is needed to smooth the model spectra.  
Differences in observing conditions can also change the instrumental resolution as a function of time,
so we calculate the line dispersion of the narrow [\mbox{O\,{\sc iii}}]$\lambda5007$ line 
for each spectrum individually and include the measurements of the line dispersion
as free parameters with Gaussian priors given by the line width measurement uncertainties.
The intrinsic narrow line width of [\mbox{O\,{\sc iii}}]$\lambda5007$ is also treated
as a free parameter with a Gaussian prior given by the line width measurement uncertainties.
For objects where the
NLR is not resolved and thus there is no intrinsic line width to the narrow line
profile, the width of the narrow emission line directly gives a measurement
of the instrumental resolution.  
Since subtracting narrow emission lines from broad emission lines can
introduce significant uncertainty into the spectrum, we model spectra
that have not had the narrow emission line subtracted and we include
the total flux of the narrow line as an additional free parameter to
be constrained by the data.

\subsection{Exploring Parameter Space}
\label{sect_dnest}

\begin{table*}
\begin{minipage}{160mm}
 \caption{BLR model parameters and their prior probability distributions.}
 \label{table_params}
 \begin{tabular}{@{}cll}
  \hline
  Parameter & Definition & Prior \\
  \hline
  $\mu$ & Mean radius of the BLR radial profile Eq.~\ref{eqn_mu} & LogUniform$(1.02\times10^{-3}\, {\rm light\,\, days}, \Delta t_{\rm data})$ \\
  $\beta$ & Unit standard deviation of BLR radial profile Eq.~\ref{eqn_beta} & Uniform$(0,2)$ \\
  $F$ & Beginning radius in units of $\mu$ of BLR radial profile Eq.~\ref{eqn_f} & Uniform$(0,1)$\\
  $\theta_i$  & Inclination angle \S~\ref{sect_angles}  & Uniform($\cos \theta_i(0,\pi/2)$) \\
  $\theta_o$  & Opening angle \S~\ref{sect_angles} & Uniform$(0,\pi/2)$ \\
  $\kappa$  & Cosine illumination function parameter Eq.~\ref{eqn_kappa}  & Uniform$(-0.5,0.5)$ \\
  $\gamma$  & Disk edge illumination parameter Eq.~\ref{eqn_gamma} & Uniform$(1,5)$ \\
  $\xi$  & Plane transparency fraction \S~\ref{sect_angular} & Uniform$(0,1)$ \\
  $M_{\rm BH}$  & Black hole mass Eq.~\ref{eqn_m} & LogUniform$(2.78\times10^4, 1.67\times10^9 M_\odot)$ \\
  $f_{\rm ellip}$  & Fraction of elliptical orbits \S~\ref{sect_inflow} & Uniform$(0,1)$ \\
  $f_{\rm flow}$  &  Flag determining inflowing or outflowing orbits \S~\ref{sect_inflow}  & Uniform$(0,1)$ \\
  $\sigma_{\rho,\,\rm circ}$  & Radial standard deviation around circular orbits \S~\ref{sect_ellip} & LogUniform$(0.001,0.1)$ \\
  $\sigma_{\Theta,\,\rm circ}$  & Angular standard deviation around circular orbits \S~\ref{sect_ellip}  &  LogUniform$(0.001,1.0)$ \\
  $\sigma_{\rho,\,\rm radial}$  & Radial standard deviation around radial orbits \S~\ref{sect_inflow} &  LogUniform$(0.001,0.1)$ \\
  $\sigma_{\Theta,\,\rm radial}$  & Angular standard deviation around radial orbits \S~\ref{sect_inflow}  &   LogUniform$(0.001,1.0)$\\
  $\sigma_{\rm turb}$  & Standard deviation of turbulent velocities \S~\ref{sect_turbulence}  &   LogUniform$(0.001,0.1)$\\
  $\theta_e$  & Angle in the $v_{\phi} - v_r$ plane \S~\ref{sect_inflow}  &   Uniform$(0,\pi/2)$\\
  \hline
 \end{tabular}
 
  \medskip
 Equation numbers refer to the first equation in which the parameter is used and section 
 numbers refer to those subsections where the parameter is defined.  
 $\Delta t_{\rm data}$ is the time span between the first and last data point in the reverberation mapping dataset.
 The prior is designated by the 
 scale in which a parameter is sampled uniformly and by the range (minimum value, maximum value).  
 Uniform$(0,1)$ denotes a uniform prior distribution between 0 and 1. 
 LogUniform$(1, 100)$ denotes a uniform prior for the log of the parameter, 
 or alternatively, a prior density $p(x) \propto 1/x$, between the parameter values $1$ and $100$.   A log-uniform prior is used for positive parameters whose order of magnitude
is unknown.
\end{minipage}
\end{table*}

\begin{figure*}
\begin{center}
\includegraphics[scale=0.67]{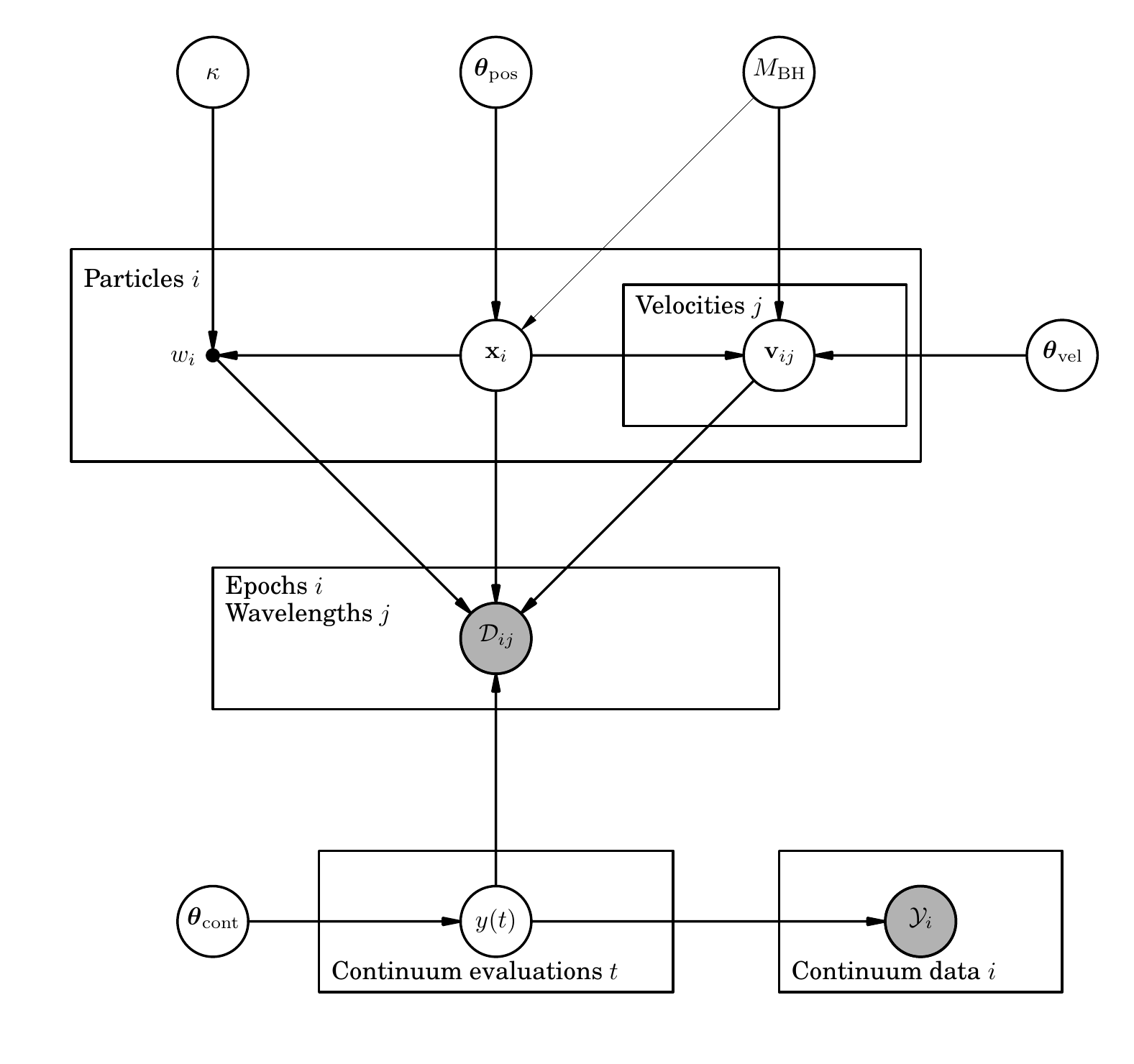}
\caption{A probabilistic graphical model 
of the parameters and their influence on simulated reverberation data created 
by the BLR model.  Each unshaded node represents a parameter (e.g. $M_{\rm BH}$) or continuum hyperparameter 
(e.g. $\boldsymbol{\theta}_{\rm cont}$) in the model and each shaded node represents a data value (e.g. $\mathcal{D}_{ij}$).  
Arrays of parameters are represented with a box, which can be thought of as a for loop.  The arrows 
represent dependence between two nodes, where the arrow between $M_{\rm BH}$ and $\mathbf{x}_i$ 
corresponds to a weak dependency due to the minimum BLR radius being set by the Schwarzschild radius.
The geometry model parameters, which determine the positions of the particles $\mathbf{x}_i$, 
include $\kappa$ and $\boldsymbol{\theta}_{\rm pos}$, a vector of the remaining geometry model parameters given in Table~\ref{table_params}: $\mu$, $\beta$, $F$, 
$\theta_i$, $\theta_o$, $\gamma$, and $\xi$. 
The dynamics model parameters, which determine the velocities of the particles $\mathbf{v}_{ij}$, include $M_{\rm BH}$ and $\boldsymbol{\theta}_{\rm vel}$,
a vector of the remaining dynamics model parameters given in Table~\ref{table_params}: $f_{\rm ellip}$, $f_{\rm flow}$, $\sigma_{\rho,\,\rm circ}$, 
$\sigma_{\Theta,\,\rm circ}$, $\sigma_{\rho,\,\rm radial}$, $\sigma_{\Theta,\,\rm radial}$, $\sigma_{\rm turb}$, and $\theta_e$.  
The continuum hyperparameters in vector $\boldsymbol{\theta}_{\rm cont}$ include $\mu_{\rm cont}$, $\sigma_{\rm cont}$, $\tau_{\rm cont}$, and $\alpha_{\rm cont}$.
This figure was made using daft-pgm.org.
} 
\label{fig_pgm}
\end{center}
\end{figure*}

Once our model of the BLR has been defined, we can explore this
high-dimensional parameter space to constrain which parameter values
best fit a specific reverberation mapping dataset by measuring the
posterior PDFs and correlations between parameters.  The full list of
all parameters in our BLR model is given in Table~\ref{table_params}
along with all of the random numbers used to assign the point particle positions
and velocities in Section~\ref{sect_model}.
 A probalistic graphical model (PGM) of the interdependence of the parameters is
shown in Figure~\ref{fig_pgm}.  One way to interpret
Figure~\ref{fig_pgm} is as a recipe for making simulated reverberation
mapping data:
\begin{enumerate}
 \item Generate a model continuum light curve using Gaussian processes and then
 \item sample it to create a realistic continuum light curve.
 \item Use BLR geometry and dynamics parameters to generate the positions and velocities of 
 all the particles in the BLR.
 \item Finally, use those positions and velocities along with the model continuum light curve to 
 make a simulated time series of broad emission line spectra or integrated broad line fluxes.
\end{enumerate}

As described in Section~\ref{sect_bayes}, we can explore
high-dimensional parameter spaces using an MCMC algorithm.
We use the diffusive nested sampling code DNest \citep{brewer11b} due to
its ability to explore correlations between parameters efficiently
 in high dimensional spaces, and
because it calculates the Bayesian evidence and thus allows for
model selection.  DNest works by using multiple walkers to explore
parameter space, starting from the prior and gradually adding hard likelihood
constraints.

 One of the inherent difficulties of fitting real data with a simplified
model is that the model is unlikely to match the data perfectly, especially
if the error bars on the data are very small. In practice, one often obtains
unrealistically precise inferences of the model parameters because the model
contains simplifications.
However we still expect the model to capture the main features of the structure
of the BLR.
  We
account for the systematic uncertainty from using a simple model by
inflating the errorbars of the data until only the macroscopic
fluctuations in the data are fit by the model.  Since we use a
Gaussian likelihood function, as discussed in
Section~\ref{sect_bayes}, we can rephrase the inflation of errorbars
as an increased weighting of the prior probability compared to the
likelihood when calculating the posterior probability.  The weighting
term is called a ``temperature" $T$, such that $\log({\rm posterior})
\propto \log({\rm prior}) + \log({\rm likelihood})/T$ and hence the
inflated errorbars are $\sqrt{T}$ larger than the original errorbars
for a Gaussian likelihood function.  Generally higher quality datasets
require larger values of $T$.  Another advantage of using nested sampling
for the computation is that we can vary the temperature and check the
sensitivity of the results without having to repeat the MCMC run.

\subsection{Limitations of the Model and Future Improvements}
\label{sect_lim}
Finally, we discuss some of the limitations of our model of the BLR
and discuss improvements to be made in the future.  One of the main
limitations of our model is the simplified dynamics of the point
particles.  We ignore the effects of radiation pressure, a force that
has a $1/r^2$ dependence like gravity, making it difficult to
disentangle from the black hole mass.  Unfortunately, this degeneracy
of radiation pressure with black hole mass means that black hole
masses could be significantly underestimated, and that the degeneracy
can only be broken by including external information about the BLR gas
density in the model \citep[][]{marconi08, marconi09, netzer09,
netzer10}.  We also ignore the self-gravity and viscosity of the BLR
gas and any interaction it has with the gas in the accretion disk.
Finally, we assume that the gas in elliptical orbits is the same gas
that could be inflowing or outflowing, when in reality the BLR could
have multiple components with different geometries and dynamics.

Another limitation to our model of the BLR is the simplified treatment
of radiative transfer, both for the ionizing and broad line photons.
We ignore any asymmetry of the ionizing photons except for an optional
preference for photons away from the BLR midplane.  We also ignore
detailed radiative transfer of line photons within the BLR gas, both
locally and globally.  While we have included two additional obscuration
effects in this new version of the BLR model, transparency of the disk
midplane to line photons ($\xi$) and asymmetry of the ionizing photons away
from the disk midplane ($\gamma$), these are simplifications of what is most
likely an inherently complicated problem.

Some of these limitations can be at least partially solved in future
models of the BLR.  For example, CLOUDY models constrain the direction
in which line photons are emitted from individual clouds of BLR gas,
as well as the emissivity and responsivity of line emission as a
function of radius \citep[e.g.][]{ferland98,ferland13}.  Using a table
of pre-computed values from CLOUDY would not only provide a more
physically-detailed local opacity to our model, but would also
constrain the radial distribution of broad line emitting gas.
 By modeling the BLR using multiple broad emission lines simultaneously, 
we can also start to place constraints on the underlying distribution
of gas density.

Recently \citet{li13} developed an independent code to model
reverberation mapping data using a geometry model of the BLR based on
the model of \citet{pancoast11}.  They include additional flexibility
in their model by allowing for non-linear response of the broad
emission lines to incident continuum radiation.  While the average
emission line response of their sample is close to linear, they find
that individual AGNs can have non-linear response, suggesting that
this effect may be important to include in future implementations of
our modeling code.

Another improvement that could be made to our model is better
treatment of the dynamics.  One option could be to include separate
geometries for each dynamical component, for example a thin disk of
gas in elliptical orbits with a cone of outflowing gas.  We could also
improve our treatment of outflows to include the detailed dynamics
found in simulations of disk winds or complex models of outflows.  For
example, instead of assuming that outflowing gas has mainly radial
trajectories at or near the escape velocity of its present position,
we could consider the more complicated case where the gas is
accelerated to velocities on the order of the escape velocity and
where the escape velocity is defined at an initial wind launching
radius instead of the current position of the gas
\citep[e.g.][]{castor75, proga99}.

Finally, breathing of the BLR may play an important role in
determining the response of emission line flux as a function of time
\citep[see][and references therein]{korista04}.  Breathing of the BLR
is where BLR emission comes from gas farther from or closer to the
central engine based on increases or decreases in the ionizing
luminosity, respectively.  If the mean radius of the BLR changes
substantially over the course of a reverberation mapping campaign,
then this could have a noticeable effect on the measured time lag
and the results from direct modeling analysis.

\section{Tests with Simulated Data and Arp~151}
\label{sect_simdata_tests}
We demonstrate the capabilities of our improved model of the BLR and direct modeling code by
recovering the model parameters for two simulated reverberation mapping datasets.  
By modeling the time series of emission line profiles using a geometry and dynamical model
of the BLR as well as modeling the integrated emission line light curve using a geometry model
of the BLR, we illustrate the benefits of a full spectroscopic dataset.

\subsection{The simulated datasets}
\label{sect_mockspectra}

\begin{table*}
\begin{minipage}{130mm}   
 \caption{Geometry and dynamics true parameter values of simulated spectral datasets and 
inferred geometry and dynamics posterior median parameter values and 68\% confidence intervals.}
 \label{table_simparams}
 \begin{tabular}{@{}ccccc}
  \hline
Geometry Model  & Simulated   & Simulated       & Simulated   & Simulated      \\
 Parameter      & Data 1 (True) & Data 1 (Inferred) & Data 2 (True) & Data 2 (Inferred) \\
  \hline
$r_{\rm mean}$ (light days) &   $ 4.0$ &   $4.19^{+0.21}_{-0.21}$ &   $ 4.0$ &   $3.54^{+0.44}_{-0.35}$ \\ 
$r_{\rm min}$ (light days) &   $ 1.0$ &   $0.85^{+0.18}_{-0.26}$ &   $ 1.0$ &   $0.89^{+0.22}_{-0.19}$ \\ 
$\sigma_{r}$ (light days) &   $ 3.0$ &   $3.23^{+0.30}_{-0.25}$ &   $ 2.4$ &   $2.39^{+0.40}_{-0.24}$ \\ 
$\tau_{\rm mean}$  (days)  &   $ 3.62$ &   $3.59^{+0.15}_{-0.15}$ &   $ 3.39$ &   $3.30^{+0.18}_{-0.15}$ \\ 
$\beta$ &   $ 1.0$ &   $0.97^{+0.09}_{-0.09}$ &   $ 0.8$ &   $0.92^{+0.09}_{-0.11}$ \\ 
$\theta_o$ (degrees) &   $40$ &   $49.0^{+ 8.4}_{- 7.6}$ &   $30$ &   $27.3^{+11.0}_{- 8.6}$ \\ 
$\theta_i$ (degrees) &   $20$ &   $20.2^{+ 2.7}_{- 3.3}$ &   $20$ &   $22.8^{+10.0}_{- 6.7}$ \\ 
$\kappa$ &   $-0.4$ &   $-0.31^{+0.09}_{-0.09}$ &   $-0.4$ &   $-0.16^{+0.31}_{-0.24}$ \\ 
$\gamma$ &   $ 5.0$ &   $2.73^{+1.29}_{-1.19}$ &   $ 5.0$ &   $3.50^{+1.02}_{-1.86}$ \\ 
$\xi$ &   $ 0.3$ &   $0.31^{+0.10}_{-0.08}$ &   $ 0.1$ &   $0.53^{+0.37}_{-0.32}$ \\ 
  \hline
Dynamical Model  & Simulated   & Simulated       & Simulated   & Simulated    \\
 Parameter      & Data 1 (True) & Data 1 (Inferred) & Data 2 (True) & Data 2 (Inferred) \\
  \hline
$\log_{10}(M_{\rm BH}/M_\odot)$ &   $ 6.5$ &   $6.42^{+0.06}_{-0.05}$ &   $ 6.5$ &   $6.48^{+0.08}_{-0.26}$ \\ 
$f_{\rm ellip}$ &   $ 0.0$ &   $0.07^{+0.05}_{-0.04}$ &   $ 1.0$ &   $0.84^{+0.12}_{-0.50}$ \\ 
$f_{\rm flow}$ &   $ 0.0$ &   $0.25^{+0.18}_{-0.17}$ &   -- &   $0.32^{+0.28}_{-0.22}$ \\ 
$\theta_e$ (degrees) &   $ 0.0$ &   $ 7.9^{+ 7.1}_{- 5.0}$ &   -- &   $70.2^{+16.7}_{-38.6}$ \\ 
$\sigma_{\rm turb}$ &   $ 0.0$ &   $0.024^{+0.055}_{-0.021}$ &   $ 0.0$ &   $0.009^{+0.026}_{-0.007}$ \\ 
  \hline
 \end{tabular}
 
  \medskip
 The columns with (True) give the true parameter values for the simulated datasets and the columns with (Inferred) give
the inferred parameter values and their uncertainties.  True parameter values with -- are unimportant for that specific
simulated dataset.
\end{minipage}
\end{table*}

\begin{figure*}
\begin{center}
\includegraphics[scale=0.55]{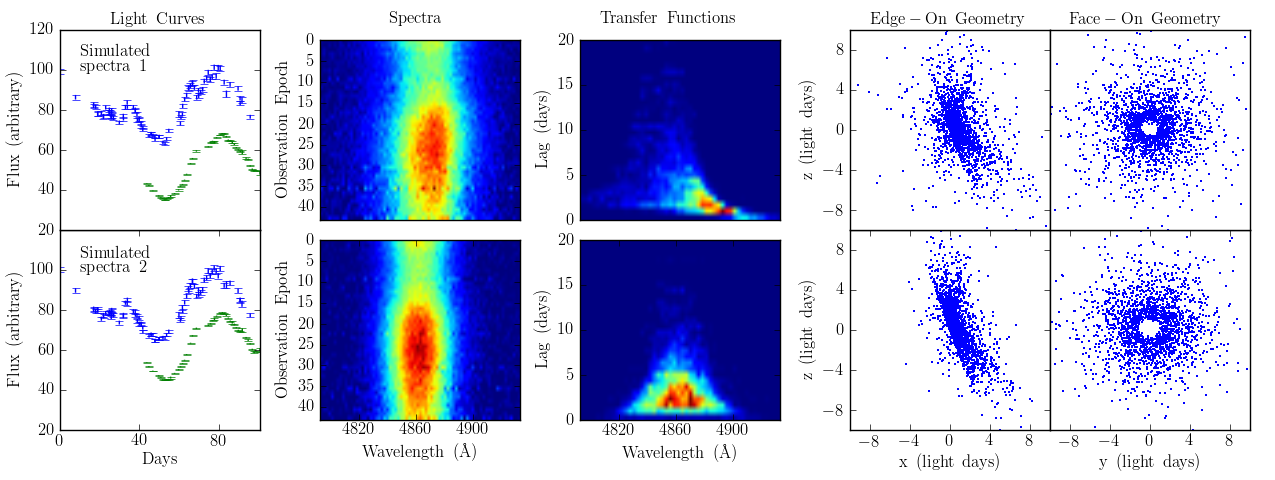}
\caption{Simulated spectral time series 1 (top row) and 2 (bottom row).  First (leftmost) 
column shows the integrated line light curve in green and the continuum light curve in blue.  
Second column shows the spectral time series over the wavelength range of the emission line 
as a function of time series epoch.  Third column shows the transfer function as a function of time lag 
and wavelength.  Fourth and fifth columns show the edge-on and face-on views, 
respectively, of the BLR geometries for each simulated dataset (the observer views the origin
from the positive x-axis).} 
\label{fig_mockspectra}
\end{center}
\end{figure*}

In order to generate realistic simulated reverberation mapping datasets, we use the LAMP 2008
dataset of \Hb\ emission for Arp 151 \citep{walsh09, bentz09} to determine the sampling cadence, flux errors, 
instrumental smoothing, and approximate scale of the BLR.  
The Arp 151 dataset includes a {\it B}-band continuum light curve and a time series of \Hb\ emission
line profiles, where the broad and narrow \Hb\ flux is isolated from the spectrum using spectral 
decomposition techniques as described by \citet{park12a}.  
As described in Section~\ref{sect_dnest}, the simulated datasets
are created by first generating a model of the Arp 151 continuum light curve using Gaussian 
processes and sampling that model continuum light curve with the same cadence as for Arp 151.
We then add Gaussian noise to the continuum light curve using the error vector of the Arp 151
light curve.  Next we set fixed the BLR geometry and dynamics model parameters to the values
found in Table~\ref{table_simparams} and sample the \Hb\ emission line profile at the times given by
the Arp 151 spectral dataset.  Finally, we add Gaussian noise to the model spectra based on the 
spectral errors in the Arp 151 dataset.  In order to account for the fact that real reverberation
mapping datasets are likely more complicated than our model of the BLR assumes, we inflate the spectral errors
and added Gaussian noise on the simulated dataset by a factor of three compared to the Arp 151 dataset, 
to obtain more realistic uncertainties on the inferred model parameters.
To reduce numerical noise in the simulated spectra to less than the uncertainty in the spectral fluxes, 
we use 2000 particles and assign each one ten independent velocity values.  
The width of the narrow line component of \Hb\ is modeled using the line dispersion of the narrow [\mbox{O\,{\sc iii}}]$\lambda5007$
emission line from the Arp 151 dataset, calculated for each epoch of spectroscopy.
The instrumental resolution is then measured by comparing the measured line dispersion for [\mbox{O\,{\sc iii}}]$\lambda5007$
with its intrinsic line width as calculated by \citet{whittle92}.

The simulated datasets are based on the geometry and dynamics inferred for the LAMP 2008
dataset in paper II as shown in Table~\ref{table_simparams}, with small mean radii for the BLR of $4$ light days, close to 
exponential radial profiles with $\beta \sim 1$, substantial width to the BLR of $\sim 2-3$ light days,
thick disks with opening angles of $30-40$ degrees, close to face-on inclination angles of 20 degrees,
preferential emission from the far side of the BLR ($\kappa = -0.4$) and the edges of the disk ($\gamma = 5$), 
and mostly opaque mid-planes ($\xi = 0.1-0.3$).  The black hole masses are also chosen to be similar
to the LAMP 2008 sample with $M_{\rm BH} = 10^{6.5} M_\odot$ and each of the simulated datasets
is dominated by either elliptical orbits or inflowing orbits.  
The differences between the simulated datasets can also 
be easily seen in Figure~\ref{fig_mockspectra}, which shows not only the continuum, 
line, and spectral timeseries, but also the transfer functions and geometries of the BLR.  
The simulated spectral datasets consist of the following:
\begin{enumerate}
 \item A thick, wide disk with an exponential profile and dynamics dominated by
 inflowing orbits.
 \item A thinner, narrower disk, with a radial profile between a Gaussian and exponential and 
 dynamics dominated by elliptical orbits.
\end{enumerate}
The continuum light curve interpolation using Gaussian
processes is also held constant for all three simulated datasets, although the random noise added
to each realistically sampled continuum light curve is different.  

 While not an issue for simulated emission line profiles, real reverberation mapping data 
must contend with ambiguity in multiple spectral components overlying the broad emission
line profile.  For example, with \Hb\ we not only have possible overlap of the red wing with the 
narrow [\mbox{O\,{\sc iii}}] emission lines and the blue wing with He\,{\sc ii}, but there may also be substantial overlap with Fe\,{\sc ii} broad line emission.  Blending between multiple broad components is especially important to
disentangle because the different broad emission lines will be generated in BLR gas at different radii from the 
black hole and blending could confuse the dynamical modeling results.  
In order to isolate any single broad emission line profile, it is 
necessary to apply a method of spectral decomposition that will remove most of the ambiguity in 
overlapping spectral components.

\subsection{Recovery of model parameters: spectral datasets}

\begin{figure*}
\begin{center}
\includegraphics[scale=0.85]{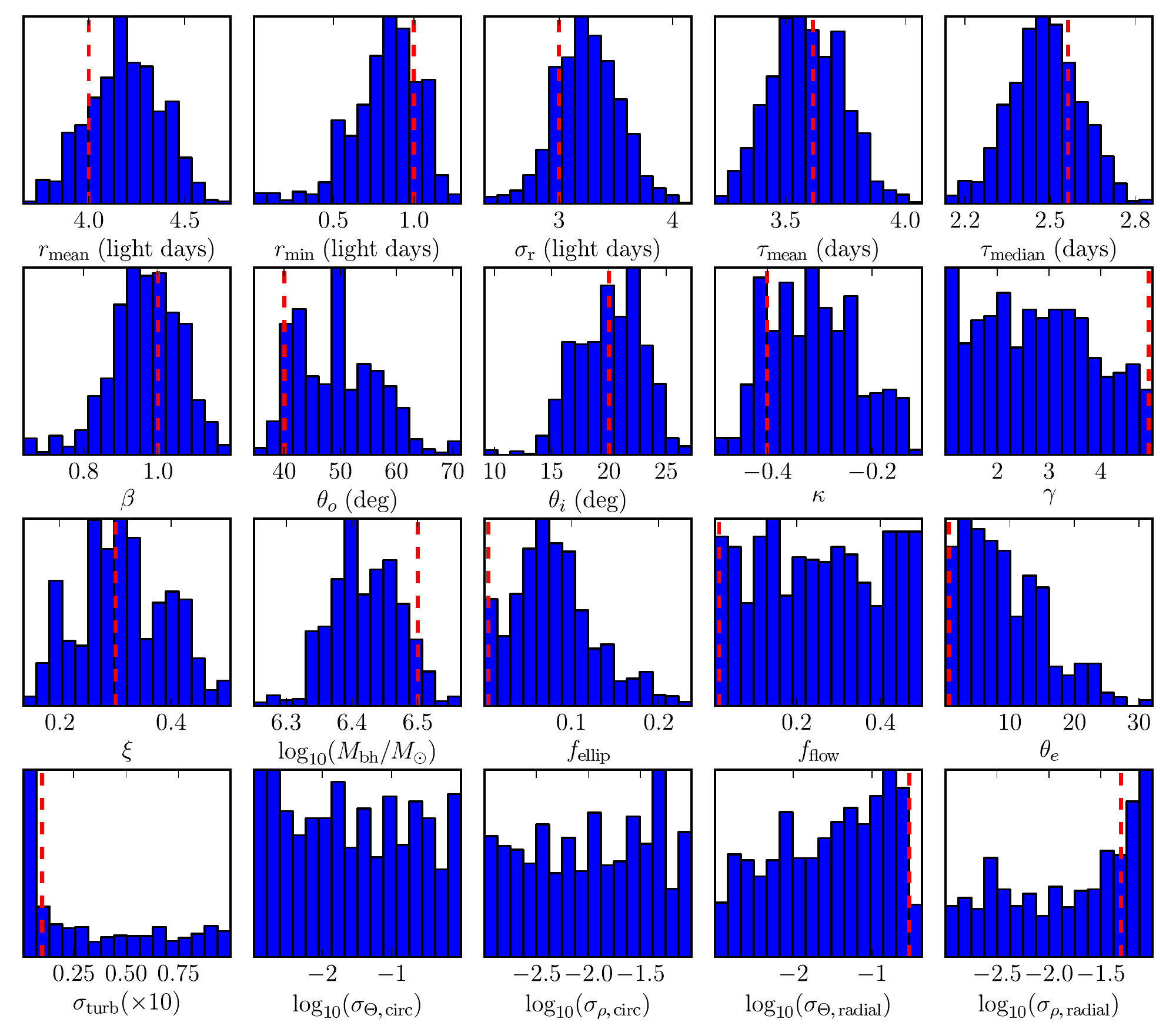}
\caption{Inferred model parameters for simulated spectral dataset 1.  The true parameter values are given by
the vertical dashed red lines for those cases where the true value affects the shape of the simulated spectral dataset.} 
\label{fig_posteriors1}
\end{center}
\end{figure*}

\begin{figure*}
\begin{center}
\includegraphics[scale=0.85]{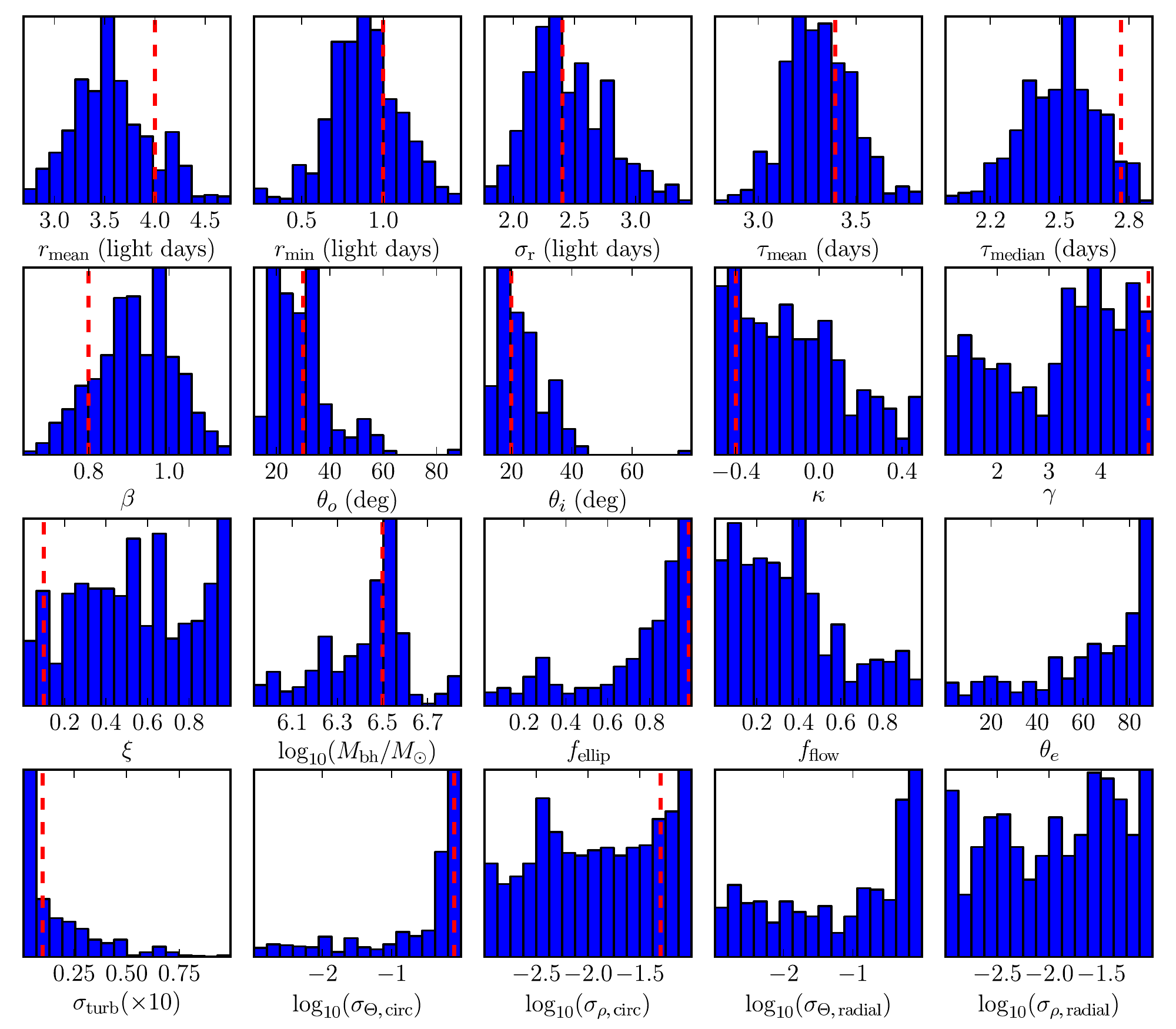}
\caption{The same as Figure~\ref{fig_posteriors1} for simulated spectral dataset 2.} 
\label{fig_posteriors2}
\end{center}
\end{figure*}

\begin{figure}
\begin{center}
\includegraphics[scale=0.35]{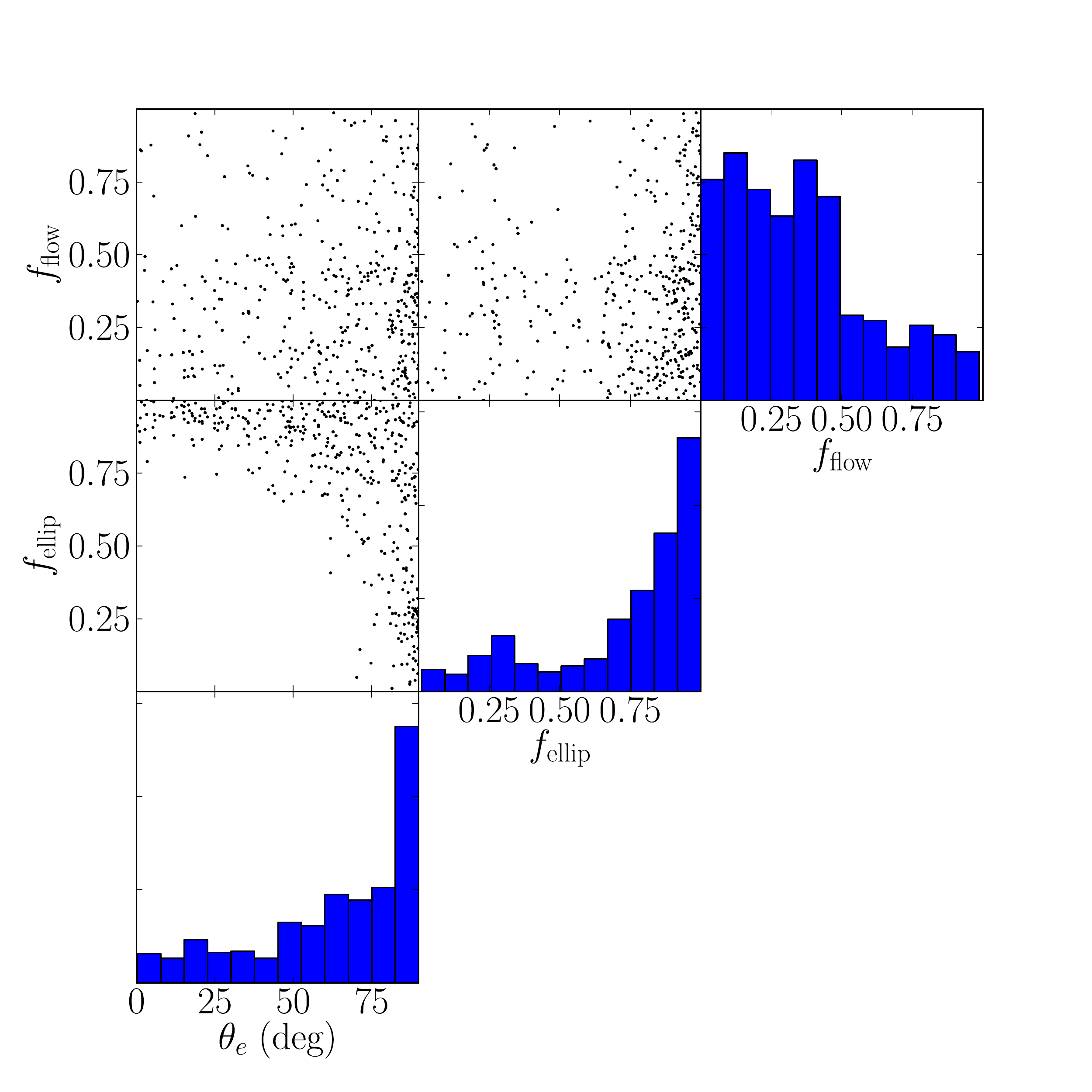}
\caption{Marginal posterior PDFs and correlations between parameters for simulated dataset 2, 
including the fraction of elliptical orbits ($f_{\rm ellip}$), the flag determining inflowing or outflowing orbits ($f_{\rm flow}$),
and the angle in the $v_\phi - v_r$ plane ($\theta_e$).} 
\label{fig_corner2}
\end{center}
\end{figure}

As a first test of our direct modeling code and BLR model, we attempt
to recover the true parameter values of the two simulated spectral
datasets described in Section~\ref{sect_mockspectra}.  We assume the
same instrumental resolutions as a function of time that are used to generate the simulated
datasets and use 2000 particles and assign ten independent
velocities to each one.  Since we add Gaussian noise to the simulated
datasets, we do not expect to recover every parameter of the BLR
exactly.  In addition, certain BLR geometries and dynamics make it difficult to
constrain certain parameters.  For example, when the majority of particles
are in elliptical orbits, the fraction of particles in inflowing or outflowing orbits
may not be well constrained.  Or, a nearly face-on very thin
disk will make it difficult to constrain the parameters $\kappa$,
$\gamma$, and $\xi$ since these parameters affect the relative line emission
throughout the height of the very thin disk in this case.

The inferred posterior PDFs for the BLR geometry and dynamics model
parameters are shown in Figures~\ref{fig_posteriors1} and
\ref{fig_posteriors2} for simulated
datasets 1 and 2, respectively.  The true parameter values for the
simulated datasets are shown by vertical red dashed lines for comparison, and
in the cases where the true parameter value does not matter (e.g. when
the dynamics are entirely dominated by elliptical orbits so there is
no true value of $f_{\rm flow}$) no red line is given.  Overall, the
modeling code is able to recover the true parameter values to within
reasonable uncertainties, as listed in
Table~\ref{table_simparams}.  
Specifically, we constrain the mean radius of the BLR to within 0.5 light days uncertainty,
the mean time lag to within 0.2 days uncertainty, and the inclination and opening angles
to within $\sim 10$ degrees.  The geometry parameters that add asymmetry to the BLR 
are more difficult to constrain, with $\kappa$ and $\xi$ well constrained for simulated dataset 1 while
neither $\kappa$, $\gamma$, nor $\xi$ are well-determined for simulated dataset 2.  

We also constrain
$\log_{10}(M_{\rm BH}/M_\odot)$ to $0.05 - 0.25$ dex uncertainty, where the variation
comes mainly from larger correlated uncertainties with the inclination and 
opening angles for simulated dataset 2.  The dynamics are also well-recovered for both
simulated datasets, with a clear preference for inflow in simulated dataset 1 and for elliptical
orbits centered around the circular orbit values in simulated dataset 2.  A clearer picture of
the preference for elliptical orbits for simulated dataset 2 can be seen in Figure~\ref{fig_corner2},
which shows the correlations between $f_{\rm ellip}$, $f_{\rm flow}$, and $\theta_e$.  
Specifically, for values of $\theta_e$ approaching 90 degrees, the distribution of inflowing or outflowing
orbits becomes identical to the distribution for elliptical orbits centered around the circular orbit value in
the $v_\phi - v_r$ plane.  This means that when $\theta_e \sim 90$ degrees, although $f_{\rm ellip}$ and $f_{\rm flow}$
are mostly unconstrained, the velocity distribution for the particles is very similar to that of $f_{\rm ellip} \sim 1$.

In general, these two simulated spectral datasets show that we can expect to 
obtain substantial constraints on the geometry and dynamics of the BLR for reverberation
mapping datasets similar in quality to LAMP 2008.  The potential constraints on the black hole mass
are also promising, although they depend upon the geometry of the BLR, specifically the precision with which 
we can measure the inclination and opening angles.

\subsection{Recovery of model parameters: integrated line datasets}
\label{sect_linetest}

For those cases where a full spectroscopic reverberation mapping
dataset is not available, we can apply a geometry-only model of the
BLR and reproduce integrated emission line flux light curves.  We test
whether this approach provides constraints on the geometry of the BLR
that are comparable to the full geometry plus dynamical modeling
problem using the simulated datasets described in
Section~\ref{sect_mockspectra} and shown in the left panel of
Figure~\ref{fig_mockspectra}.

We find that the mean time lag and mean radius are well constrained
with geometry-only modeling.  The mean and median time lag inferred for each
simulated dataset are given in Table~\ref{table_lag_compare}, along
with the true mean and median lag values and the value measured by CCF analysis.
The inferred mean time lag is not only accurate, but the inferred
uncertainty in the mean time lag through geometry-only modeling of
$\sim 0.25$ days is almost half as large as for the CCF time lag.
This shows that geometry-only modeling is a promising tool for
measuring time lags.  The mean radius is inferred with slightly larger
uncertainties to be $3.58^{+1.18}_{-0.97}$ light days for simulated
dataset 1 and $2.90^{+0.97}_{-0.24}$ for simulated dataset 2, while the true mean
radius is $4$ light days.  

Unfortunately the other geometry model parameters are not as well
constrained.  The parameters $\gamma$ and $\xi$ are completely
unconstrained for both of the simulated datasets, and $\theta_o$,
$\theta_i$, $\beta$, $F$, and $\kappa$ are mostly unconstrained.
Generally there is a slight preference for a specific value of
$\theta_o$, $\theta_i$, $\beta$, $F$, and $\kappa$, but none or almost
none of the parameter space is ruled out.  These results for
geometry-only modeling suggest that a full spectroscopic reverberation
mapping dataset is needed to constrain the geometry of the BLR, since
otherwise there are too many degeneracies between model parameters to
infer anything other than the mean time lag and mean radius
consistently.

\subsection{Comparison with JAVELIN}

\begin{table}
 \caption{Comparison of BLR geometry modeling, JAVELIN, and CCF lag measurements.}
 \label{table_lag_compare}
 \begin{tabular}{@{}ccc}
  \hline
  Lag (days)                    & Sim Data 1                      & Sim Data 2 \\
  \hline
  True mean lag                       &  $3.62$                              & $3.39$  \\
   True median lag                   &  $2.56$                              & $2.77$  \\
  $\tau_{\rm mean}$        &  $3.36^{+0.20}_{-0.15}$  &  $3.29^{+0.23}_{-0.17}$ \\
  $\tau_{\rm median}$      &  $2.61^{+0.25}_{-0.21}$  &  $3.10^{+0.17}_{-0.18}$ \\
  $\tau_{\rm JAVELIN}$    &  $2.94^{+0.13}_{-0.12}$  &  $3.21^{+0.13}_{-0.14}$ \\
  $\tau_{\rm cen}$            &  $3.70^{+0.50}_{-0.48}$  &   $3.62^{+0.56}_{-0.40}$\\
    \hline
 \end{tabular}

  \medskip
  $\tau_{\rm mean}$ and $\tau_{\rm median}$ are the mean and median time lags inferred from BLR geometry modeling, $\tau_{\rm JAVELIN}$ is the time 
  lag measured by JAVELIN, and $\tau_{\rm cen}$ is the center-of-mass lag measured from the CCF.
\end{table}

Recently another method has been developed for measuring the time lag
in reverberation mapping data using integrated emission line light
curves by \citet{zu11}.  This method has been implemented in an
open-source code called JAVELIN written in Python.\footnote{Download
JAVELIN here: https://bitbucket.org/nye17/javelin} JAVELIN works by
using a top-hat transfer function with two parameters, a mean lag and
a width of the top hat.  The continuum light curve in JAVELN is
interpolated using a CAR(1) model, which is equivalent to the
continuum model implemented here.  The parameter space of the
continuum light curve and transfer function models is sampled using
MCMC, providing posterior PDFs for the model parameter values.

We can compare recovery of the time lag using BLR geometry modeling of
integrated emission line light curves to the results from JAVELIN.
For simulated dataset 1, we measure a mean lag of $\tau_{\rm JAVELIN}
= 2.94^{+0.13}_{-0.12}$ days and a mean width of the top-hat transfer function of
$w = 7.33^{+0.26}_{-0.30}$ days using JAVELIN.  This can be compared to
the true mean lag of 3.62 days and the true median lag of 2.56 days for simulated dataset 1 to see that
the mean lag measured by JAVELIN is between the true mean and median lags. 
For simulated dataset 2, we measure $\tau_{\rm
JAVELIN} =  3.21^{+0.13}_{-0.14}$ days and $w =  5.26^{+0.82}_{-0.63}$ days.  
Again, the mean lag measured by JAVELIN is between the true mean lag of 3.39 days
and the true median time lag of 2.77 days, although closer to the mean lag.  
The tendency for the time lag measured by JAVELIN to fall closer to the true
mean or median lag is due to the shape of the transfer function; 
in very asymmetric transfer functions, the mean and median time lag are 
increasingly discrepant, with JAVELIN more sensitive to the true median
time lag for very asymmetric transfer functions.
  
While the tendency of JAVELIN to measure a time lag 
ranging between the true mean and median time lags 
may appear to complicate its interpretation, 
an uncertainty of $\sim 1$ day from the difference 
between the true mean and median lags is comparable to the
uncertainty introduced by additional assumptions, as discussed
in Section~\ref{sect_ccf_tests}, when using time lag measurements to estimate
the mean radius of the BLR or to measure the black hole mass.
These comparisons suggests that JAVELIN is a excellent resource for 
measuring the time lag even if the JAVELIN lag uncertainties do not 
reflect the uncertainty introduced by asymmetric transfer functions. 
However, to constrain more than the time lag,
more flexible modeling of the transfer function must be done.
  
In comparison, the CCF lag measurements for the two simulated datasets
agree with the true mean lag values due to larger uncertainties.  The CCF
lag measurements do not agree more closely with the true median lag values than
with the true mean lag values for more asymmetric transfer functions, as for JAVELIN lags. 
The quoted error bars on the CCF lag values, $\tau_{\rm cen}$, in 
Table~\ref{table_lag_compare} are calculated by drawing 
a random subset of the line and continuum light curves points, with the
same number of random draws as the original light curves.  For points in 
the light curves that are drawn $N$ times, the flux error is reduced by $\sqrt{N}$. 
Finally, the randomly drawn light curve fluxes are modified by adding 
random Gaussian noise given by the flux errors.  
 This is similar to the ``flux randomization"/``random subset selection" (FR/RSS) 
approach described in \citet{peterson98} except
the FR/RSS approach throws out any redundant points in the light curve instead
of reducing the flux errors by $\sqrt{N}$, resulting in slightly larger 
uncertainties in the CCF lag.
The CCF time lag is measured for 1000 iterations of this sequence and
we quote the median and 68\% confidence intervals of the CCF time lag distributions.
For the two simulated
datasets tested here with data quality comparable to the LAMP 2008 dataset for
Arp 151 \citep{bentz09}, the error bars are $\sim 0.5$ days, or $\sim 14$\% the
value of $\tau_{\rm cen}$.  
This comparison suggests that while CCF analysis may not
give the most precise measurement of the mean or median time lag, 
the CCF lag uncertainties likely include much of 
the systematic uncertainties from an unknown transfer function. 

 Finally, we also consider the effects of detrending the light curves
before calculating the CCF lag.  Detrending can improve
the shape of the CCF when there are strong long-term trends that can
be removed by subtracting a linear fit to the light curves \citep{welsh99}.
Since our simulated data do not contain strong long-term trends,
detrending the light curves should have minimal impact on the 
measured CCF lags.  We confirm this by subtracting a linear fit
to the simulated continuum light curves from both the continuum 
and line light curves.  Due to the difference in length between the 
continuum and line light curves, fitting the continuum and line light
curves with linear fits separately destroys the correlation between
the light curves.  When we use a linear fit to the continuum light curve
to detrend both light curves we obtain CCF lag measurements 
for simulated dataset 1 of $\tau_{\rm cen} = 3.37^{+0.48}_{-0.37}$ days
and for simulated dataset 2 of $\tau_{\rm cen} = 3.38^{+0.47}_{-0.37}$ days,
which agree to within the uncertainties with the un-detrended CCF lag values.

\subsection{Dynamical modeling without a full spectral dataset}
As shown in Section~\ref{sect_linetest}, a spectroscopic dataset
offers substantially more information about the BLR for direct
modeling.  Here we explore an intermediate case, where the available
data consist of the usual continuum light curve, an integrated
emission line light curve, and a mean spectrum.  Since the mean
spectrum contains some information about the kinematics of the BLR, we
can model this dataset using the fully dynamical model of the BLR.
However, with only the mean spectrum, this dataset cannot constrain
the time lag as a function of velocity or wavelength, as possible for a full
spectroscopic dataset.

In order to provide a test of this intermediate dataset case that is
as realistic as possible, we use the LAMP 2008 dataset for Arp 151.  A
description of the dataset can be found in paper II.  In the analysis
of this test, we focus on the differences in inferred parameter values
between this test and the full dynamical modeling results presented in
paper II.  In general, the modeling results for the full spectroscopic
dataset and for the intermediate dataset are fully consistent, but the
uncertainty on the inferred model parameter values is much larger for
the intermediate dataset.  For example, the black hole mass is inferred
to have a posterior PDF with a long tail at high masses, giving 
$\log_{10}(M_{\rm BH}/M_\odot) = 6.74^{+0.66}_{-0.13}$ compared
to the value from Paper II of $\log_{10}(M_{\rm BH}/M_\odot) = 6.62^{+0.10}_{-0.13}$.
Similarly, the uncertainty in $\theta_i$, $\theta_o$, and $\kappa$ is larger by at least 
a factor of 3, the uncertainty in $\xi$ is larger by at least
a factor of two, and $\gamma$ is completely
undetermined for the intermediate dataset.  The two marginally
consistent results are the mean radius and mean lag, which are both
substantially larger for the intermediate dataset and have
uncertainties at least 10 times larger than for the full
spectroscopic dataset.  This is due to a preference for $\beta \to 2$, corresponding
to heavy-tailed radial distributions where the median radius and median lag 
are more consistent measurements of the characteristic size of the BLR.
Overall, this test suggests that considerable
information about the BLR can be inferred from the mean line profile,
but the constraints on BLR geometry and dynamics parameters are
significantly better when the full spectroscopic dataset is used.
Finally, while this intermediate dataset allows for measurement of the
black hole mass, it cannot be constrained to less than the $\sim
0.4$ dex scatter in the $f$ factor due to a tail in the posterior PDF at high masses.

\section{Comparison with cross-correlation analysis}
\label{sect_ccf_tests}
We can compare the results of direct modeling to the standard
reverberation mapping analysis of using the cross-correlation function
(CCF) to measure time lags and the dispersion or FWHM of the broad
emission line to measure a characteristic velocity of the BLR.  In
addition to providing a sanity check on our direct modeling results,
such a comparison allows us to explore some of the uncertainties
involved in standard reverberation mapping analysis.  First, we
consider the time lag traditionally measured using the CCF, how it
compares to a measurement of the mean radius and how sampling of the
line light curve and variability of the continuum light curve affect
its measurement.  Second, we consider the combination of the CCF lag
with measurements of the emission line width to explore the
uncertainty in black hole masses measured using the virial product.

\subsection{Comparing the time lag and mean radius}
\label{sect_rmean_lag}

\begin{figure}
\begin{center}
\includegraphics[scale=0.5]{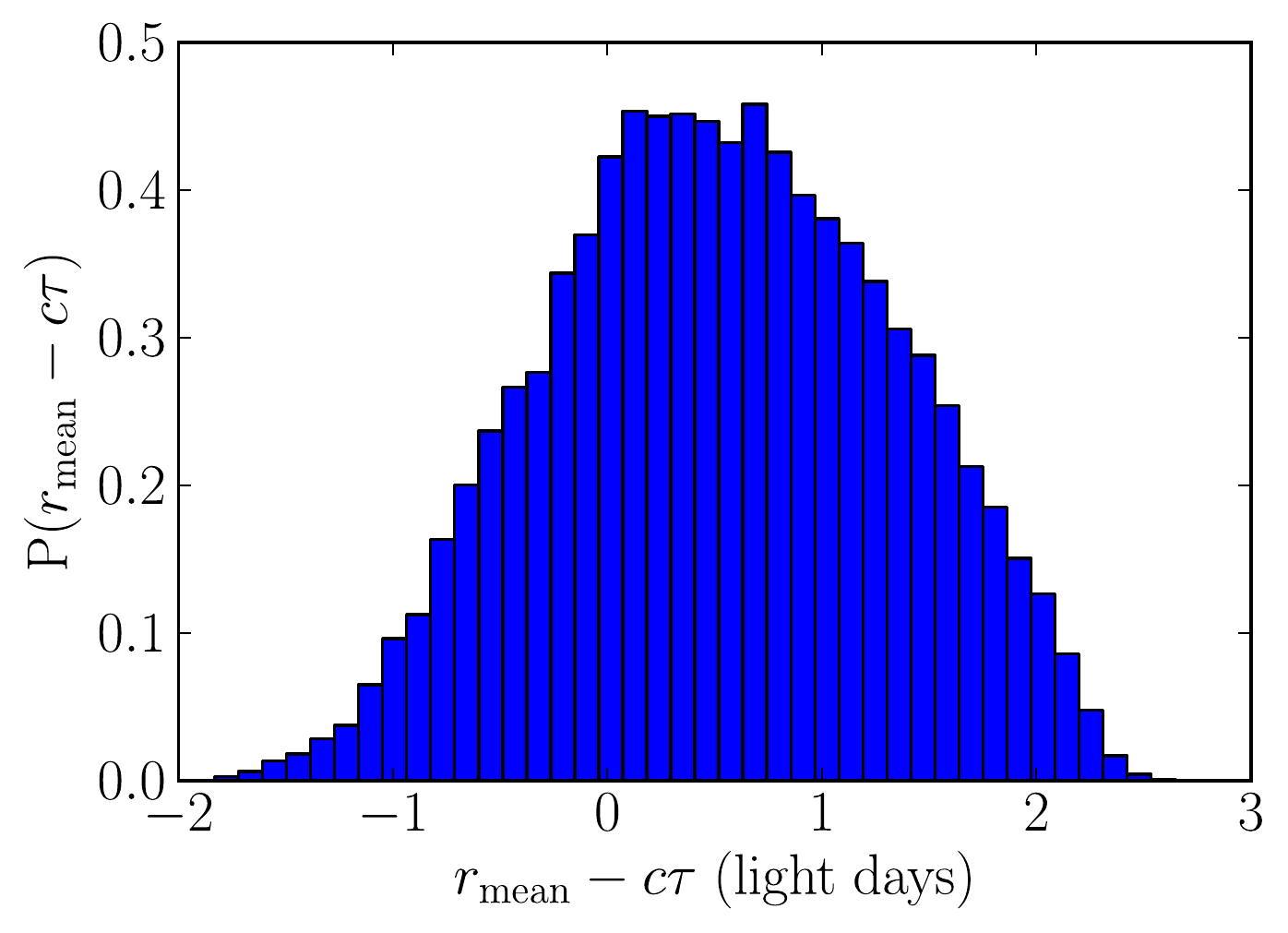}
\caption{Difference between the mean radius and mean lag for BLR models drawn randomly 
from the prior with $\mu = 4$ light days.  The distribution is asymmetric because the parameter 
$\xi$, the BLR plane transparency fraction, only shortens the mean lag compared to the mean 
radius, creating an excess of models where the mean radius is larger than the mean lag.} 
\label{fig_radius_lag_test}
\end{center}
\end{figure}

One of the main assumptions in the traditional analysis is that the
time lag measured from CCF analysis is related to some characteristic
radius of the BLR.  We explore the validity of this assumption by
comparing the mean radius and the mean time lag in our geometry model
of the BLR.  We hold the mean radius fixed at $\mu = 4$ light days and
allow the other geometry model parameter values to sample their priors
as listed in Table~\ref{table_params}, with the exception of the
inclination angle, which we constrain to vary between zero (face-on)
and 45 degrees.  The results of this comparison are shown in
Figure~\ref{fig_radius_lag_test} for 200,000 samples.  The difference
between the mean radius, $r_{\rm mean}$, and the mean lag, $\tau$, is
generally greater than one, meaning that the mean lag (in days) is
usually shorter than the mean radius (in light days).  This is due to
the geometry parameter $\xi$ that allows the midplane of the BLR to be
transparent or opaque, since a BLR midplane that is not transparent
will result in fewer particles with longer lags and hence a
tendency for the mean lag to be smaller than the mean radius.  The
mean of $r_{\rm mean} - \tau$ is 0.53 light days and the standard
deviation of the distribution is 0.80 light days.  This suggests that
the uncertainty in using the time lag as a measurement of the mean
radius is relatively small, on the order of the CCF time lag
uncertainty typically quoted for high-quality reverberation mapping
data of $\sim 1$ day.

\subsection{The effects of line light curve sampling}
\label{sect_sampling_test}

\begin{table}
 \caption{Geometry model parameter values of simulated emission line light curves used in the comparison of 
 direct modeling with the cross-correlation analysis approach.}
 \label{table_geo_mod_true}
 \begin{tabular}{@{}cccccccc}
  \hline
  Mock & $\tau_{\rm mean}$ & $\theta_i$ &  $\theta_o$ &  $\kappa$  & $\beta$ &  $F$   &  $\xi$\\
  Line   & (days) &  (deg) & (deg)  &         &         &   &  \\
  \hline
  1 &  3.69  &  10 & 25 & -0.25 & 1.0 & 0.2    &  0.5 \\
  2 &  3.77  &  10 & 25 & 0.5 & 0.11 & 0.5    &  1  \\
  3 &  4.01  &  10 & 90 & 0.0 & 0.11 & 0.99   &  1  \\
  4 &  5.34  &  10 & 90 & -0.5 & 0.11 & 0.99   &  1  \\
  5 &  4.00  &  0 & 0.5 & 0.0 & 0.11 & 0.99     &  1 \\
  \hline
 \end{tabular}

  \medskip
  All simulated datasets were created with a mean radius, $\mu$, of 4 light days and 
  with $\gamma = 1$.
\end{table}

\begin{figure}
\begin{center}
\includegraphics[scale=0.6]{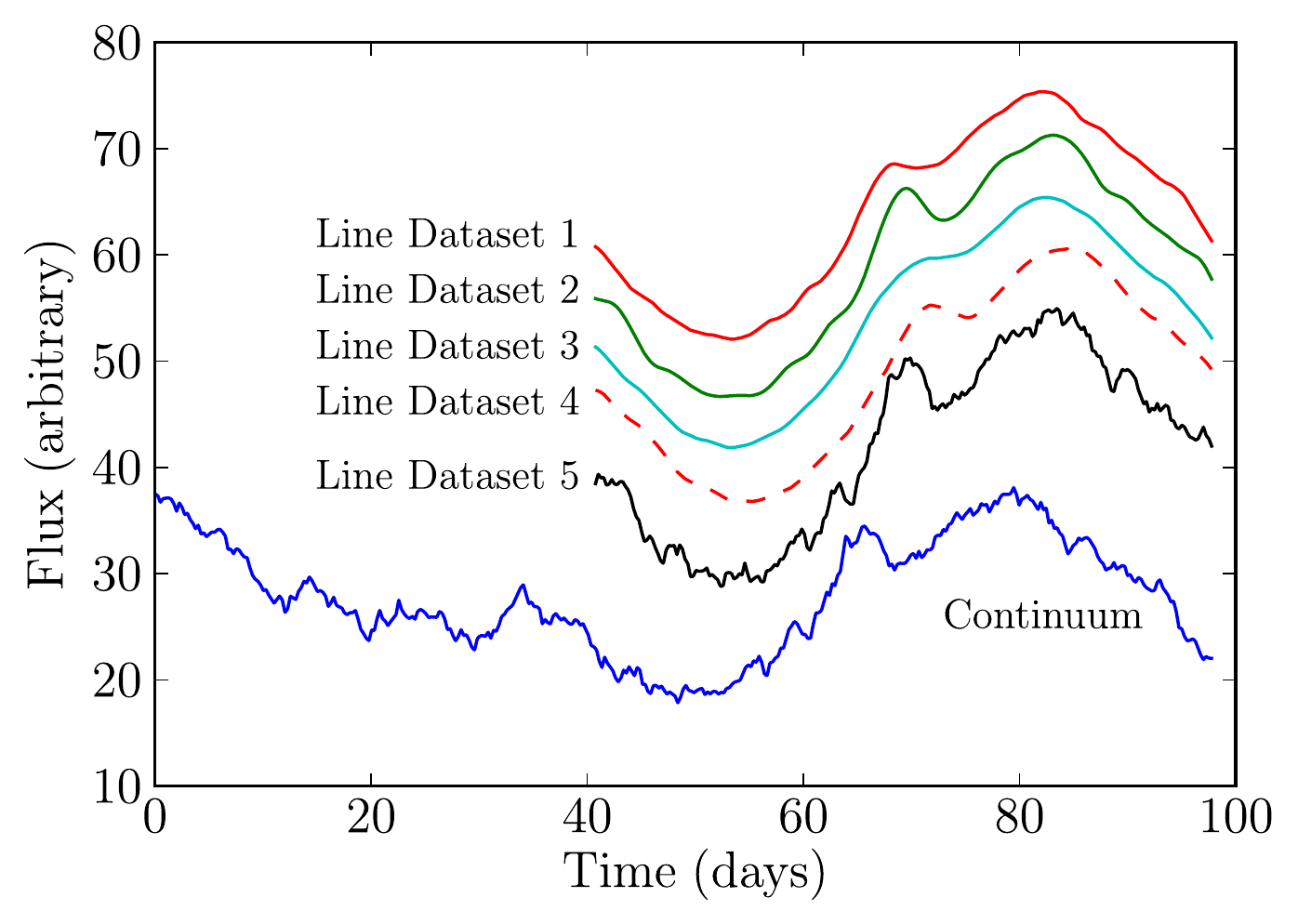}
\caption{Simulated integrated emission line datasets including the AGN continuum light curve in solid 
blue and integrated \Hb\ line light curves in solid red, green, cyan, black, and dashed red.  The 
continuum light curve is based on the LAMP 2008 light curve of Arp 151 \citep{walsh09}.  The 
simulated \Hb\ line light curves correspond to five different BLR geometries, as shown in Figure \ref{fig_mockgeo}. } 
\label{fig_mocklc}
\end{center}
\end{figure}

\begin{figure}
\begin{center}
\includegraphics[scale=0.62]{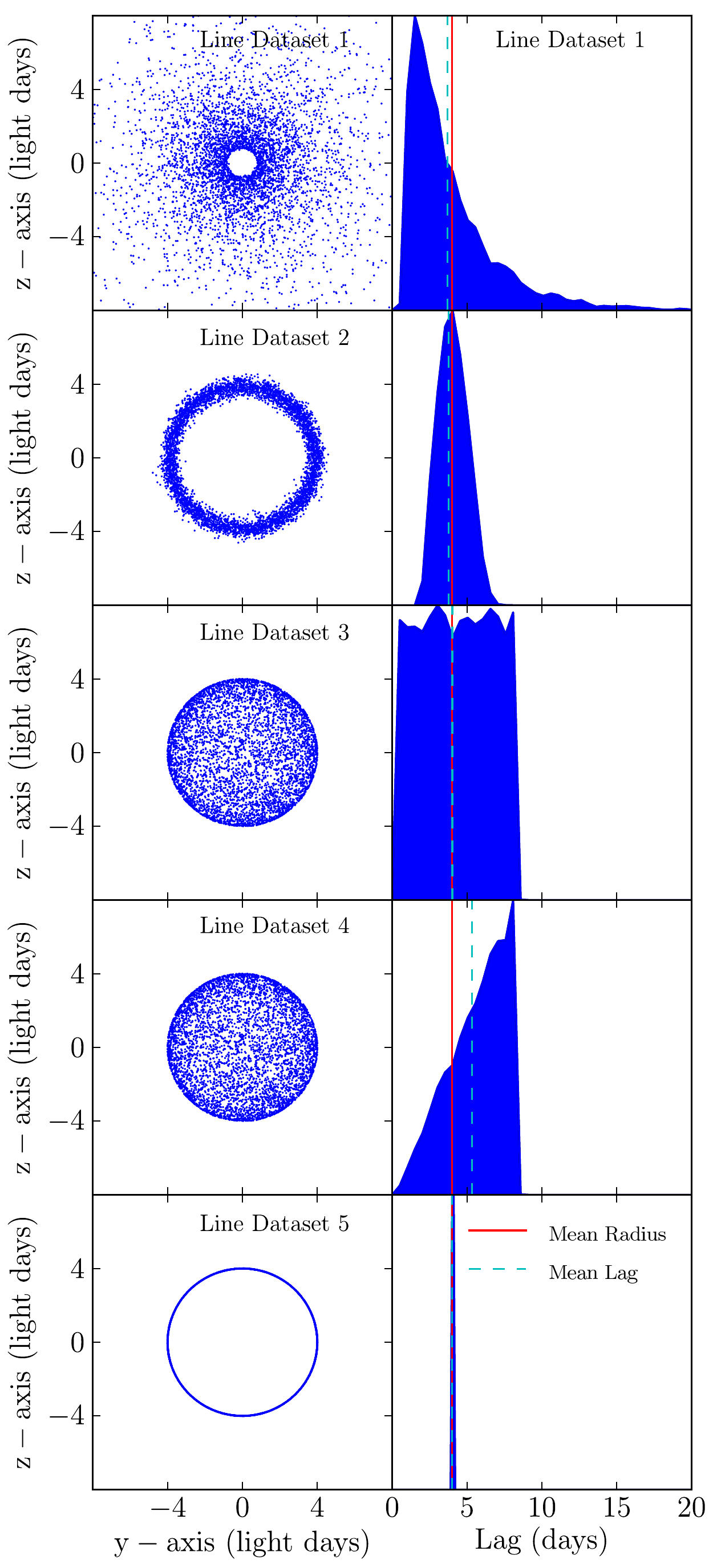}
\caption{Geometries of the BLR (left panels) and corresponding transfer functions (right panels) 
of the simulated reverberation mapping datasets shown in Figure \ref{fig_mocklc}. Top to bottom BLR 
geometries: face-on wide disk, face-on donut, spherical shell, spherical shell with preferential 
emission from the back of the sphere, and a face-on thin ring. }      
\label{fig_mockgeo}
\end{center}
\end{figure}

\begin{figure}
\begin{center}
\includegraphics[scale=0.54]{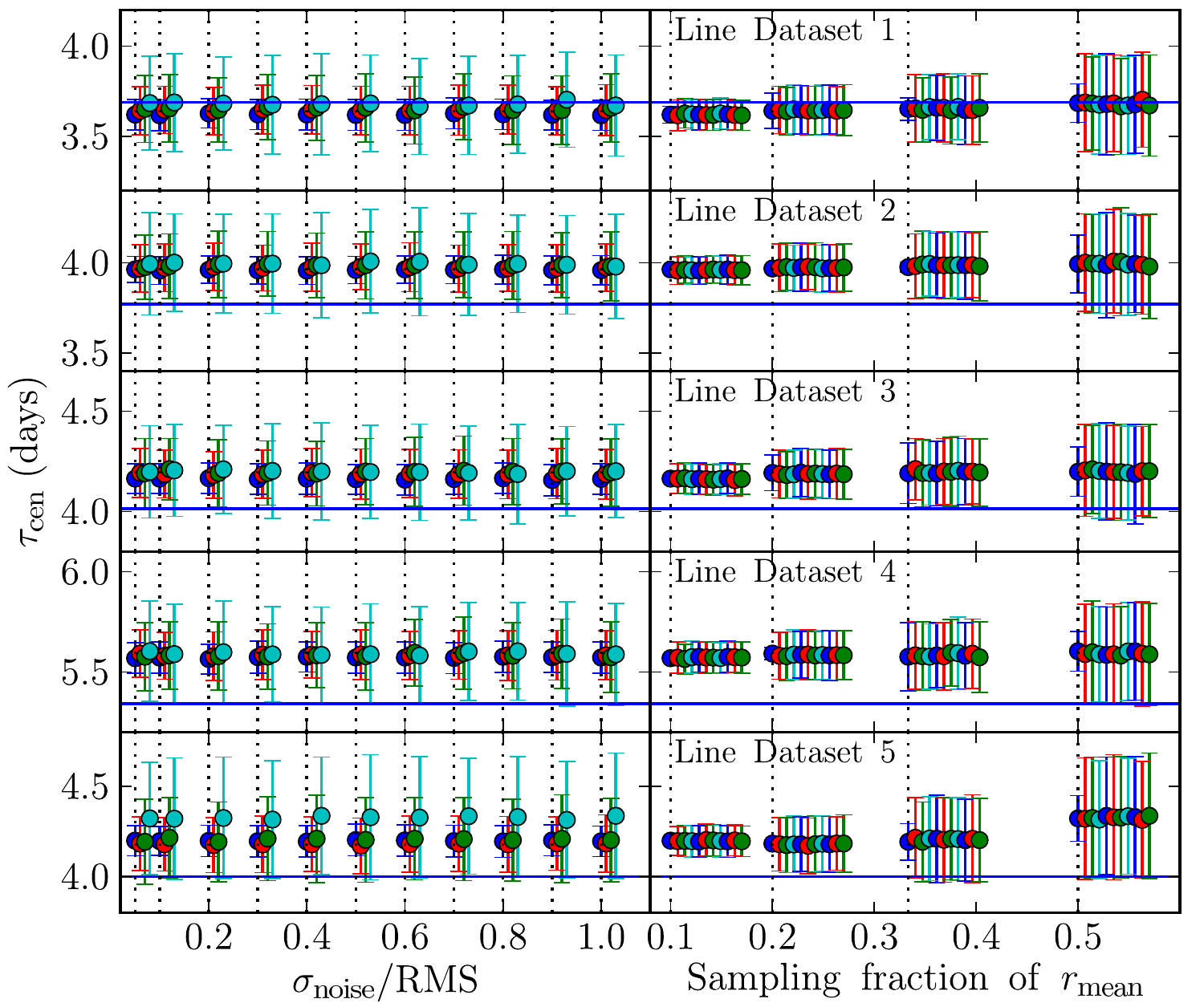}
\caption{The CCF lag $\tau_{\rm cen}$ as a function of the ratio of $\sigma_{\rm noise}$ to
the RMS variability of the line light curve and versus the cadence as given by the sampling
fraction of the mean radius.  Top to bottom: simulated line dataset 1, 2, 3, 4, and 5.  
The horizontal blue lines show the true mean time lags for each simulated line dataset.  The 
vertical dotted lines show the values of the x-axis for which each cluster of points correspond,
where the clusters of points are spread out along the x-axis to show their spread.  
These results are for the case where 3/4 of the epochs are not lost to weather.} 
\label{fig_line_sampling2}
\end{center}
\end{figure}

Next we explore the dependence of the measured CCF lag 
on the geometry of the BLR and on the
sampling characteristics of the emission line light curve.
We focus on four very simple BLR geometries and one more
realistic one, as shown in Figure~\ref{fig_mockgeo}, including
\begin{enumerate}
  \item A nearly face-on wide disk with preferential emission from the far side and a
  disk midplane that is more than half opaque.
  \item A nearly face-on ring with preferential emission from the near side.
  \item A spherical shell (making a top-hat transfer function).
  \item A spherical shell with preferential emission from the far side.
  \item A perfectly face-on thin ring (making a $\delta$-function transfer function).
\end{enumerate}
We use these five geometries of the BLR to create simulated emission
line light curves, as shown in Figure~\ref{fig_mocklc} using the same
input continuum light curve and with very fine sampling of 0.1 days
for both the line and continuum light curves.  The geometry model
parameter values are given in Table~\ref{table_geo_mod_true}.  In
order to test how the quality of integrated emission line light curves
affects measurement of the CCF lag, we degraded the quality of the
simulated data by adding random Gaussian noise to the line light curve
and by reducing the sampling cadence.  For each simulated dataset
degraded by adding $\sigma_{\rm noise}$ of random Gaussian noise, by
sampling the line light curve at some fraction of the true mean radius
of the BLR, and by losing a fraction of that sampled line light curve
to weather, we computed the CCF lags $\tau_{\rm cen}$ and $\tau_{\rm
peak}$ for 1000 realizations of assigning the random noise and losing
a fraction of the light curve to weather.  The simulated line light
curves were degraded by:
\begin{enumerate}
 \item Reducing the sampling cadence to 1/10, 1/5, 1/3, or 1/2 of the true 
 mean radius value of the geometry model of 4 light days.  This means that the highest 
 cadence is about half a day.  
 \item Adding random Gaussian noise, $\sigma_{\rm noise}$, at the level of 0.05, 0.1, 0.2, 0.3, 
 0.4, 0.5, 0.6, 0.7, 0.8, 0.9, or 1.0 times the RMS variability of the simulated line light curve.
 \item Including only a fraction of the total number of line light curve data points randomly 
 from the light curve to simulate observations lost to weather.  The fractions are 1, 3/4, 2/3, and 1/2.    
\end{enumerate}

Some illustrative results of this comparison are shown in
Figure~\ref{fig_line_sampling2}, with the lefthand column showing the
CCF lag $\tau_{\rm cen}$ versus the ratio of $\sigma_{\rm noise}$ over
the RMS variability and with the righthand column showing the CCF lag
$\tau_{\rm cen}$ versus the cadence as a sampling fraction of $r_{\rm
mean}$.  Figure~\ref{fig_line_sampling2} shows the results for when
3/4 of the line light curve is not lost to weather.  The trend
continues for larger fractions of the light curve lost to weather: the
uncertainties on the measured CCF lag $\tau_{\rm cen}$ increase while
the mean lag measurement stays the same.  For the case where no
observations are lost to weather, the error bars become comparable to
the size of the points in Figure~\ref{fig_line_sampling2}.  
Overall,
these results suggest that for different geometries of the BLR
$\tau_{\rm cen}$ can be offset from the true lag value by as much as a
quarter of a light day (for a true mean lag of $\sim 4$ light days,
see Table~\ref{table_geo_mod_true} for the exact values), which is
well within typical uncertainties on CCF time lags quoted in the
literature.  For light curves with larger flux errors and lower
cadence, this offset is easily within the error bars.  In addition to
a possible offset from the true lag values, these results show the
importance of sampling the light curve at smaller fractions of the
mean lag, even when the signal to noise quality of the light curve is
high.  As the fraction of the light curve lost to weather increases,
this effect becomes more important.
 Detrending of the simulated light curves does not change these
results.

\subsection{The effects of continuum variability}
\label{sect_continuum_test}

\begin{figure}
\begin{center}
\includegraphics[scale=0.55]{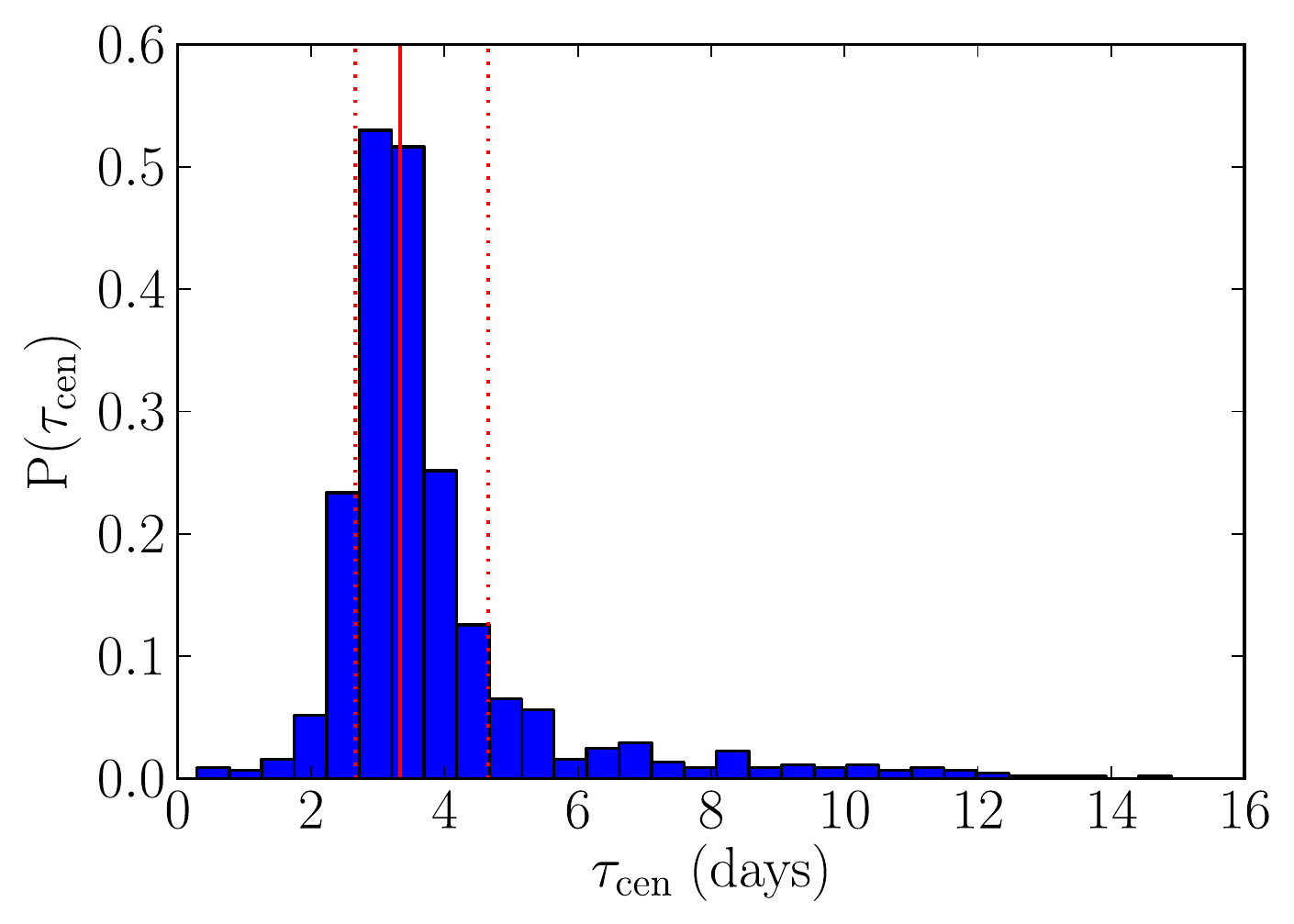}
\caption{Histogram of the CCF center-of-mass time 
lag $\tau_{\rm cen}$ for 1000 random continuum light curve model realizations.  
The true mean lag is $3.74$ days.
The vertical solid red line denotes the median value of 3.33 days and the 
dotted vertical red lines give the 68\% confidence interval around the median at 
2.66 and 4.66 days.} 
\label{fig_continuum_test}
\end{center}
\end{figure}

In addition to light curve sampling effects, there is also the
possibility that variability features in the AGN continuum light curve
could affect measurement of CCF lags.  We explore this source of
uncertainty by generating 1000 random realizations of AGN continuum
light curves, keeping the continuum hyper-parameters fixed to values
similar to those inferred for Arp 151 and the BLR geometry model fixed
to the values for simulated integrated line dataset 1 given in
Table~\ref{table_geo_mod_true}.  Given each realization of the AGN
continuum light curve and the fixed BLR geometry model, we generate an
integrated emission line light curve.  We use the sampling cadence of
the LAMP 2008 dataset for Arp 151, described in
Section~\ref{sect_mockspectra}, for each realization of the continuum
and line light curves.  We then calculate the CCF center-of-mass lag
$\tau_{\rm cen}$ for each of the 1000 realizations, obtaining
successful CCF lag measurements for over 90\% of the random continuum
realizations.

The results are shown in Figure~\ref{fig_continuum_test} as a
histogram of $\tau_{\rm cen}$ values, where we have truncated the
histogram to between zero and fifteen days for clarity.  The median
and 68\% confidence interval for all measurements of $\tau_{\rm cen}$
is $3.33^{+1.33}_{-0.67}$ days, and considering only values of
$\tau_{\rm cen}$ between zero and fifteen days reduces the
uncertainties by less than 0.1 days.  
 Detrending of the simulated light curves results in a similar median
value for $\tau_{\rm cen}$ of $3.23^{+0.97}_{-0.61}$ days.
These inferred median values for
$\tau_{\rm cen}$ agree to within the uncertainties with each other and the true
value of the mean lag of $3.74$ days.  This test demonstrates that the
main consequence of continuum variability is to add additional scatter
to measurements of $\tau_{\rm cen}$ on the order of $\sim 1$ day,
without shifting the median measurement of $\tau_{\rm cen}$ away from
the true value.

\subsection{Comparing the black hole mass and virial product}

\begin{figure}
\begin{center}
\includegraphics[scale=0.55]{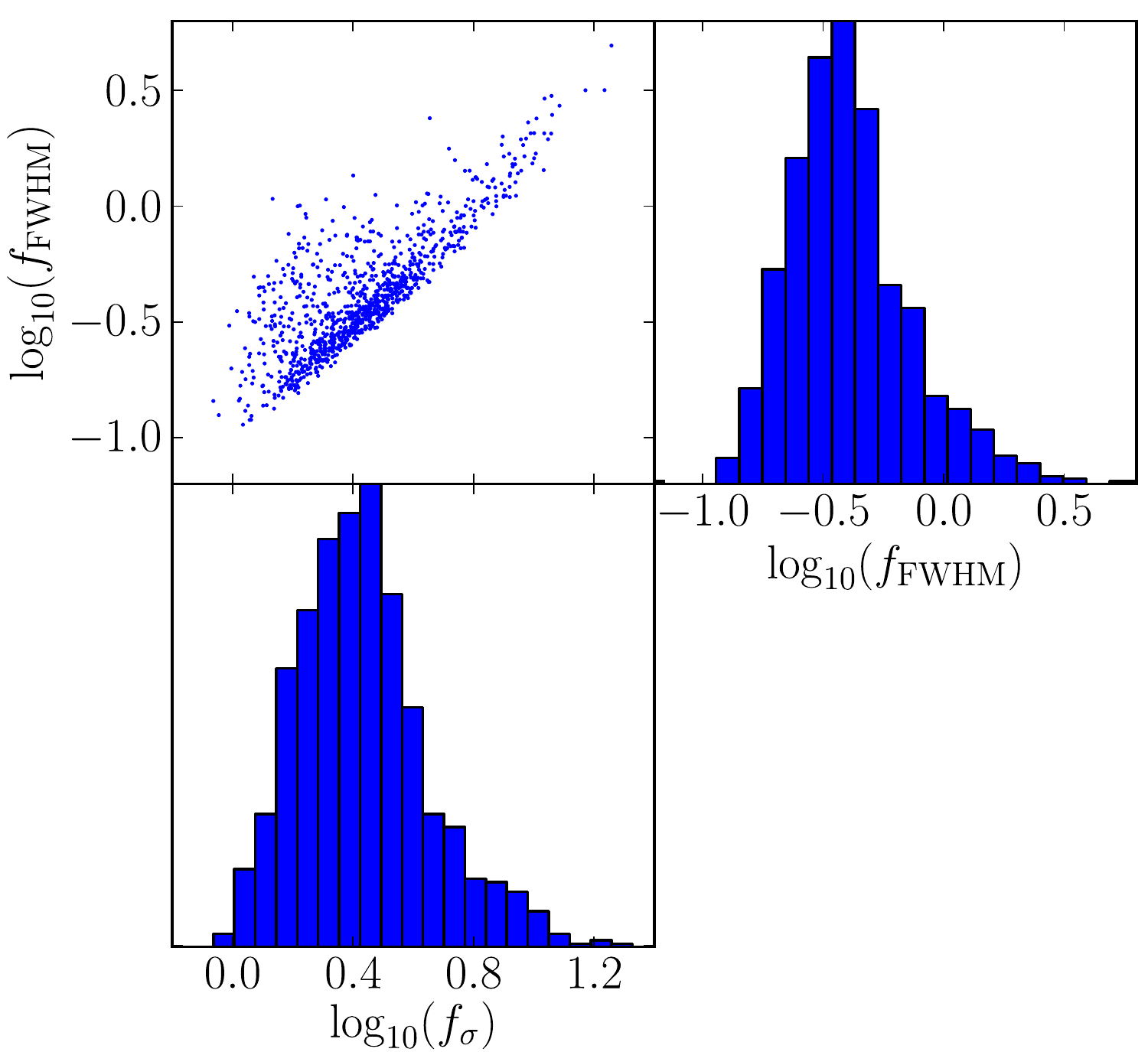}
\caption{Distributions of $f$ factor values for a fixed value of black hole mass
and mean radius, with the other BLR model parameters allowed to vary.  $f_\sigma$
and $f_{\rm FWHM}$ are calculated using the CCF lag $\tau_{\rm cen}$ and
 the line dispersion of the RMS emission
line profile or the FWHM of the mean emission line profile respectively.  The top
left panel shows the correlation between $f_\sigma$ and $f_{\rm FWHM}$.} 
\label{fig_f_compare}
\end{center}
\end{figure} 

Other than the mean lag from CCF analysis, the black hole mass
measured from the virial product is the key measurement of
reverberation mapping studies.  However the use of the virial product,
$M_{\rm vir} = c\tau \Delta v^2/G$, to measure black hole mass involves
making many assumptions, including that the mean lag is a good measure
of the physical scale of the BLR and that the width of the broad
emission line is a good measure of the velocity field of the BLR.  We
attempt to quantify the uncertainty introduced by these assumptions by
calculating the virial product from instances of our geometry and
dynamics BLR model.  We hold the black hole mass fixed at $M =
10^{6.5}M_\odot$ and the mean radius fixed at $\mu = 4$ light days
while allowing all other geometry and dynamics model parameters to
vary within their prior bounds, except for the inclination angle, which
is limited to within zero (face-on) and 45 degrees.  For each sample
of a BLR model, we calculate the CCF lag $\tau_{\rm cen}$, the line
dispersion of the RMS emission line profile, and the full width at
half maximum (FWHM) of the mean line profile.  Using these three
values we can calculate the virial product either using the line
dispersion or FWHM line width measurement.  By further dividing the
true black hole mass by the virial product we can work in terms of the
virial coefficient $f$, where $f_\sigma$ is calculated from the virial
product using the line dispersion and $f_{\rm FWHM}$ is calculated
from the virial product using the FWHM.

The results for the comparison of true black hole mass to virial
product are shown in Figure~\ref{fig_f_compare} for 1000 samples of
the BLR model parameters (other than $M$ and $\mu$, which are held
fixed).  The cadence of the continuum light curve and spectral time series
were based on the cadence of the LAMP 2008 dataset for Arp 151,
as described by \citet{bentz09}.
The values of $\log_{10}(f_\sigma)$ and $\log_{10}(f_{\rm
FWHM})$ are clearly correlated, but there is significant scatter in
the relation.  More importantly, the dispersion in the $f$ factors is
encouragingly small: the mean value of $\log_{10}(f_\sigma)$ is 0.43,
with a standard deviation of 0.22, and the mean value of
$\log_{10}(f_{\rm FWHM})$ is $-0.39$ with a standard deviation of 0.26,
 where the dispersion in the $f$ factors does not depend on whether
the light curves have been detrended. 
This means that if the BLR is well described by our phenomenological
model, we should not be surprised that the $M_{\rm BH} - \sigma_*$
relation based on reverberation mapping black hole mass measurements
does not have much larger scatter than the canonical $\sim 0.4$ dex
found for galaxies with dynamical mass measurements.

\section{Conclusions}
\label{sect_conclusions}
In this paper we present an improved and expanded simply parameterized phenomenological
model of the BLR for direct modeling of reverberation mapping data.  In addition to 
describing the model in detail, we test the performance of the direct modeling approach
using simulated reverberation mapping datasets with and without full spectral information.
We also use this model of the BLR to explore sources of uncertainty in the traditional
cross-correlation analysis used to measure time lags in reverberation mapped AGNs
as well as sources of uncertainty in traditional measurements of the black hole mass
using the virial product.  Our main conclusions are as follows:
\begin{enumerate}
\item For simulated data with the same properties as the LAMP 2008 spectroscopic dataset for
Arp 151, we can recover the black hole mass to within 0.05-0.25 dex uncertainty and
distinguish between elliptical orbits and inflow.  We recover the mean radius
and mean lag with $5-12$\% uncertainties  and the opening angle of the disk
and inclination angle to within $5-10$ degrees.
\item For the same simulated datasets, but where integrated emission line fluxes
are used instead of the full spectroscopic information, we can use a BLR geometry model
to constrain the mean radius
and mean lag with $5-35$\% uncertainties and obtain only minimal constraints 
on the geometry of the BLR. 
\item Using a combination of an integrated emission line light curve and a mean emission 
line profile for direct modeling allows for some constraints on the geometry of the BLR, but
with greater uncertainty than from using the full spectroscopic dataset.  The uncertainty
in $\log_{10}(M_{\rm BH}/M_\odot)$ is also greater compared to using
the full spectroscopic dataset.
\item Comparison of BLR geometry modeling results to those from JAVELIN \citep{zu11}
and CCF analysis shows that JAVELIN recovers a time lag between
the true mean and median lag, while CCF analysis recovers a time lag closer to the true mean lag.  While
the larger lag uncertainties from CCF analysis may reflect the unknown shape of the transfer
function, the lag uncertainties from JAVELIN are smaller than the difference between the
true mean and median time lag.  
\item By considering the range in possible BLR geometries of our model, we estimate 
the uncertainty in converting a mean lag into a mean radius to be $\sim 25$\%.
\item The CCF lag $\tau_{\rm cen}$ can be offset from the true lag of a BLR model 
depending on the geometry.  Both signal-to-noise of the flux light curve and sampling
rate affect the dispersion in how far the CCF lag is relative to the true lag.  Gaps in
the light curve due to weather also introduce more uncertainty in the CCF lag.
\item For a given BLR geometry, changes in the variability features of the AGN 
continuum light curve introduces an uncertainty of $\sim 25$\% into measurements
of the CCF lag $\tau_{\rm cen}$.
\item By considering the range in possible BLR geometries and dynamics of our model,
we estimate the uncertainty in measuring the black hole mass using the virial product
to be smaller than the spread in the $M_{\rm BH} - \sigma_*$ relation.  We find that the
standard deviation of $f = M_{\rm BH}/M_{\rm vir}$ is only $\sim 0.25$ dex, i.e. smaller than the uncertainty typically quoted for virial mass estimates.
\end{enumerate}

The tests presented here demonstrate the unique capabilities
of dynamical modeling of reverberation mapping data to constrain the geometric and kinematic properties
of the BLR.  While we can use hybrid datasets consisting of integrated line flux measurements and 
a mean emission line profile, considerably more information is available from modeling the reverberations
across the emission line profile.  The improvements we have made to this simply parameterized
phenomenological model of the BLR have increased the flexibility of the method to fit a wider variety
of emission line profiles.  Future improvements will add a deeper connection to photoionization physics,
relating the distribution of broad line emission to the distribution of underlying gas, and explore the
effects of non-gravitational forces, important for inferring the correct black hole mass.

These tests also confirm that the uncertainties inherent in the traditional analysis of 
measuring lags using the cross-correlation function and black hole masses using
the virial product are relatively small, although larger than the formal uncertainties.
The simplified problem of modeling integrated emission line light curves using
a geometry-only model for the BLR presents an alternative approach for measuring
time lags and mean radii of the BLR compared to the traditional analysis.  One 
advantage to measuring time lags and mean radii with geometry modeling of the BLR
is that the final uncertainties reflect the unknown underlying transfer function.


\section*{Acknowledgements}  
We would like to thank Aaron Barth, Mike Goad, Keith Horne, Daniel Proga, and Ying Zu for helpful discussions. 
AP acknowledges support from the NSF through the Graduate Research Fellowship Program. 
AP, BJB, and TT acknowledge support from the Packard Foundation through a Packard Fellowship to TT and support from the NSF through awards NSF-CAREER-0642621 and NSF-1108835.    
BJB is partially supported by the Marsden Fund (Royal Society of New Zealand).


\appendix
\section{Comparison of dynamics model to previous work}
In previous versions of our geometric and dynamical model of the BLR, we used a less general dynamical model in which point particle 
velocities were drawn from distributions of energy $E$ and angular momentum $L$ centered around 
the circular orbit energy and angular momentum values $E_{\rm circ}$ and 
$L_{\rm circ}$ \citep[e.g.][]{pancoast11, brewer11, pancoast12}. 
We can solve for $E_{\rm circ}$ by evaluating Equation \ref{eqn_m} when $v_r = 0$ and $v_\phi = v_{\rm circ}$:
\begin{equation}
 E_{\rm circ} = -\frac{1}{2}\frac{GM}{r}.
\end{equation} 
Similarly, we can solve Equation~\ref{eqn_m} for $L$, setting $v_r=0$ and plugging in $E_{\rm circ}$ 
to find an expression for the angular momentum of a particle in a circular orbit:
\begin{equation}
 L_{\rm circ} = \sqrt{rGM}.
\end{equation}
This previous E/L model incorporated inflow and outflow in the BLR gas through the fraction of 
elliptical orbits with positive or negative radial velocity component solutions.  The inflowing and 
outflowing gas in the E/L model is thus always bound to the black hole. 

\begin{figure}
\begin{center}
\includegraphics[scale=0.44]{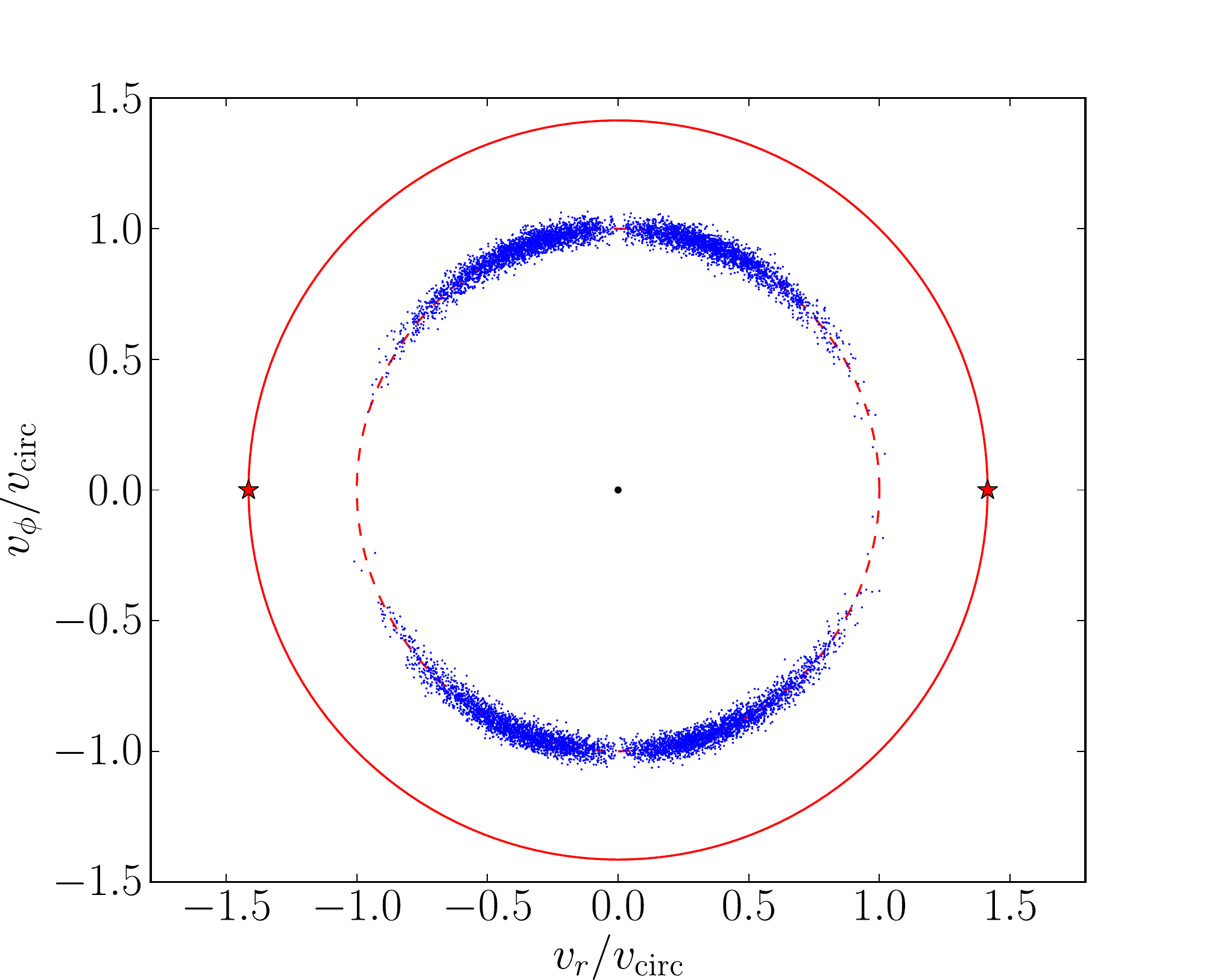}
\caption{Distributions of radial and tangential velocities, $v_r$ and $v_\phi$ for the previous 
E/L model.  Blue points are particle velocities drawn from distributions in energy and 
angular momentum centered around the point for circular orbits $(v_r, v_\phi) = (0, v_{\rm circ})$.  
Outflow corresponds to positive $v_r$.  The outer solid red circle at a radius of $|v| = \sqrt{2GM/r}$ 
denotes the velocity beyond which orbits are unbound.  The red dashed circle at a radius of  
$|v| = \sqrt{GM/r}$ denotes velocities with magnitude of the circular velocity and the circle 
around which bound elliptical orbits were distributed in the E/L model.} 
\label{fig_EL}
\end{center}
\end{figure}

For comparison to the current more general dynamical model in Figure~\ref{fig_EL}, 
we show the distributions of energy and
angular momentum in the $v_r - v_\phi$ plane for direct comparison
with Figure~\ref{fig_newVmodel}.  The radial and tangential velocity
distributions for the E/L model are centered around the red dashed
ring of radius $v_{\rm circ}$ and constrained to lie within the solid
red circle, corresponding to orbits that are bound to the black hole.
The velocity at which a particle becomes unbound is given by
setting $E=0$ and solving for $|v| = \sqrt{v_r^2 + v_\phi^2} =
\sqrt{2GM/r}$, which is $\sqrt{2}v_{\rm circ}$.  Unfortunately, while
the distributions of $E$ and $L$ are centered around their circular
orbit values, the chance of having a particle draw a close to circular
orbit is vanishingly small, as shown by the lack of points in Figure
\ref{fig_EL} at $(v_r, v_\phi) = (0, v_{\rm circ})$.  This suggests a
better way of including the circular orbit solution: draw point
particle velocities directly from distributions in radial and
tangential velocity space instead of in $E$ and $L$ space.

\bibliographystyle{apj}
\bibliography{references}

\begin{thebibliography}{}
\expandafter\ifx\csname natexlab\endcsname\relax\def\natexlab#1{#1}\fi

\bibitem[{{Antonucci}(1993)}]{antonucci93}
{Antonucci}, R. 1993, \araa, 31, 473

\bibitem[{{Barth} {et~al.}(2011){Barth}, {Nguyen}, {Malkan}, {Filippenko},
  {Li}, {Gorjian}, {Joner}, {Bennert}, {Botyanszki}, {Cenko}, {Childress},
  {Choi}, {Comerford}, {Cucciara}, {da Silva}, {Duch{\^e}ne}, {Fumagalli},
  {Ganeshalingam}, {Gates}, {Gerke}, {Griffith}, {Harris}, {Hintz}, {Hsiao},
  {Kandrashoff}, {Keel}, {Kirkman}, {Kleiser}, {Laney}, {Lee}, {Lopez}, {Lowe},
  {Moody}, {Morton}, {Nierenberg}, {Nugent}, {Pancoast}, {Rex}, {Rich},
  {Silverman}, {Smith}, {Sonnenfeld}, {Suzuki}, {Tytler}, {Walsh}, {Woo},
  {Yang}, \& {Zeisse}}]{barth11}
{Barth}, A.~J., {Nguyen}, M.~L., {Malkan}, M.~A., {et~al.} 2011, \apj, 732, 121

\bibitem[{{Bentz} {et~al.}(2006){Bentz}, {Peterson}, {Pogge}, {Vestergaard}, \&
  {Onken}}]{bentz06}
{Bentz}, M.~C., {Peterson}, B.~M., {Pogge}, R.~W., {Vestergaard}, M., \&
  {Onken}, C.~A. 2006, \apj, 644, 133

\bibitem[{{Bentz} {et~al.}(2009){Bentz}, {Walsh}, {Barth}, {Baliber},
  {Bennert}, {Canalizo}, {Filippenko}, {Ganeshalingam}, {Gates}, {Greene},
  {Hidas}, {Hiner}, {Lee}, {Li}, {Malkan}, {Minezaki}, {Sakata}, {Serduke},
  {Silverman}, {Steele}, {Stern}, {Street}, {Thornton}, {Treu}, {Wang}, {Woo},
  \& {Yoshii}}]{bentz09}
{Bentz}, M.~C., {Walsh}, J.~L., {Barth}, A.~J., {et~al.} 2009, \apj, 705, 199

\bibitem[{{Bentz} {et~al.}(2013){Bentz}, {Denney}, {Grier}, {Barth},
  {Peterson}, {Vestergaard}, {Bennert}, {Canalizo}, {De Rosa}, {Filippenko},
  {Gates}, {Greene}, {Li}, {Malkan}, {Pogge}, {Stern}, {Treu}, \&
  {Woo}}]{bentz13}
{Bentz}, M.~C., {Denney}, K.~D., {Grier}, C.~J., {et~al.} 2013, \apj, 767, 149

\bibitem[{{Blandford} \& {McKee}(1982)}]{blandford82}
{Blandford}, R.~D., \& {McKee}, C.~F. 1982, \apj, 255, 419

\bibitem[{{Brewer} {et~al.}(2011{\natexlab{a}}){Brewer}, {P{\'a}rtay}, \&
  {Cs{\'a}nyi}}]{brewer11}
{Brewer}, B.~J., {P{\'a}rtay}, L.~B., \& {Cs{\'a}nyi}, G. 2011{\natexlab{a}},
  Statistics and Computing, 21, 649, astrophysics Source Code Library

\bibitem[{{Brewer} {et~al.}(2011{\natexlab{b}}){Brewer}, {Treu}, {Pancoast},
  {Barth}, {Bennert}, {Bentz}, {Filippenko}, {Greene}, {Malkan}, \&
  {Woo}}]{brewer11b}
{Brewer}, B.~J., {Treu}, T., {Pancoast}, A., {et~al.} 2011{\natexlab{b}},
  \apjl, 733, L33

\bibitem[{{Castor} {et~al.}(1975){Castor}, {Abbott}, \& {Klein}}]{castor75}
{Castor}, J.~I., {Abbott}, D.~C., \& {Klein}, R.~I. 1975, \apj, 195, 157

\bibitem[{{Collier} {et~al.}(1998){Collier}, {Horne}, {Kaspi}, {Netzer},
  {Peterson}, {Wanders}, {Alexander}, {Bertram}, {Comastri}, {Gaskell},
  {Malkov}, {Maoz}, {Mignoli}, {Pogge}, {Pronik}, {Sergeev}, {Snedden},
  {Stirpe}, {Bochkarev}, {Burenkov}, {Shapovalova}, \& {Wagner}}]{collier98}
{Collier}, S.~J., {Horne}, K., {Kaspi}, S., {et~al.} 1998, \apj, 500, 162

\bibitem[{{Collin} {et~al.}(2006){Collin}, {Kawaguchi}, {Peterson}, \&
  {Vestergaard}}]{collin06}
{Collin}, S., {Kawaguchi}, T., {Peterson}, B.~M., \& {Vestergaard}, M. 2006,
  \aap, 456, 75

\bibitem[{{Denney} {et~al.}(2010){Denney}, {Peterson}, {Pogge}, {Adair},
  {Atlee}, {Au-Yong}, {Bentz}, {Bird}, {Brokofsky}, {Chisholm}, {Comins},
  {Dietrich}, {Doroshenko}, {Eastman}, {Efimov}, {Ewald}, {Ferbey}, {Gaskell},
  {Hedrick}, {Jackson}, {Klimanov}, {Klimek}, {Kruse}, {Lad{\'e}route}, {Lamb},
  {Leighly}, {Minezaki}, {Nazarov}, {Onken}, {Petersen}, {Peterson},
  {Poindexter}, {Sakata}, {Schlesinger}, {Sergeev}, {Skolski}, {Stieglitz},
  {Tobin}, {Unterborn}, {Vestergaard}, {Watkins}, {Watson}, \&
  {Yoshii}}]{denney10}
{Denney}, K.~D., {Peterson}, B.~M., {Pogge}, R.~W., {et~al.} 2010, \apj, 721,
  715

\bibitem[{{Ferland} {et~al.}(1998){Ferland}, {Korista}, {Verner}, {Ferguson},
  {Kingdon}, \& {Verner}}]{ferland98}
{Ferland}, G.~J., {Korista}, K.~T., {Verner}, D.~A., {et~al.} 1998, \pasp, 110,
  761

\bibitem[{{Ferland} {et~al.}(2013){Ferland}, {Porter}, {van Hoof}, {Williams},
  {Abel}, {Lykins}, {Shaw}, {Henney}, \& {Stancil}}]{ferland13}
{Ferland}, G.~J., {Porter}, R.~L., {van Hoof}, P.~A.~M., {et~al.} 2013, \rmxaa,
  49, 137

\bibitem[{{Goad} {et~al.}(2012){Goad}, {Korista}, \& {Ruff}}]{goad12}
{Goad}, M.~R., {Korista}, K.~T., \& {Ruff}, A.~J. 2012, \mnras, 426, 3086

\bibitem[{{Graham} {et~al.}(2011){Graham}, {Onken}, {Athanassoula}, \&
  {Combes}}]{graham11}
{Graham}, A.~W., {Onken}, C.~A., {Athanassoula}, E., \& {Combes}, F. 2011,
  \mnras, 412, 2211

\bibitem[{{Greene} {et~al.}(2010){Greene}, {Peng}, {Kim}, {Kuo}, {Braatz},
  {Impellizzeri}, {Condon}, {Lo}, {Henkel}, \& {Reid}}]{greene10b}
{Greene}, J.~E., {Peng}, C.~Y., {Kim}, M., {et~al.} 2010, \apj, 721, 26

\bibitem[{{Grier} {et~al.}(2013{\natexlab{a}}){Grier}, {Martini}, {Watson},
  {Peterson}, {Bentz}, {Dasyra}, {Dietrich}, {Ferrarese}, {Pogge}, \&
  {Zu}}]{grier13b}
{Grier}, C.~J., {Martini}, P., {Watson}, L.~C., {et~al.} 2013{\natexlab{a}},
  \apj, 773, 90

\bibitem[{{Grier} {et~al.}(2013{\natexlab{b}}){Grier}, {Peterson}, {Horne},
  {Bentz}, {Pogge}, {Denney}, {De Rosa}, {Martini}, {Kochanek}, {Zu},
  {Shappee}, {Siverd}, {Beatty}, {Sergeev}, {Kaspi}, {Araya Salvo}, {Bird},
  {Bord}, {Borman}, {Che}, {Chen}, {Cohen}, {Dietrich}, {Doroshenko}, {Efimov},
  {Free}, {Ginsburg}, {Henderson}, {King}, {Mogren}, {Molina}, {Mosquera},
  {Nazarov}, {Okhmat}, {Pejcha}, {Rafter}, {Shields}, {Skowron}, {Szczygiel},
  {Valluri}, \& {van Saders}}]{grier13a}
{Grier}, C.~J., {Peterson}, B.~M., {Horne}, K., {et~al.} 2013{\natexlab{b}},
  \apj, 764, 47

\bibitem[{{Horne}(1994)}]{horne94}
{Horne}, K. 1994, in Astronomical Society of the Pacific Conference Series,
  Vol.~69, Reverberation Mapping of the Broad-Line Region in Active Galactic
  Nuclei, ed. P.~M. {Gondhalekar}, K.~{Horne}, \& B.~M. {Peterson}, 23--25

\bibitem[{{Horne} {et~al.}(1991){Horne}, {Welsh}, \& {Peterson}}]{horne91}
{Horne}, K., {Welsh}, W.~F., \& {Peterson}, B.~M. 1991, \apjl, 367, L5

\bibitem[{{Kaspi} {et~al.}(2000){Kaspi}, {Smith}, {Netzer}, {Maoz}, {Jannuzi},
  \& {Giveon}}]{kaspi00}
{Kaspi}, S., {Smith}, P.~S., {Netzer}, H., {et~al.} 2000, \apj, 533, 631

\bibitem[{{Kelly} {et~al.}(2009){Kelly}, {Bechtold}, \&
  {Siemiginowska}}]{kelly09}
{Kelly}, B.~C., {Bechtold}, J., \& {Siemiginowska}, A. 2009, \apj, 698, 895

\bibitem[{{Korista} \& {Goad}(2004)}]{korista04}
{Korista}, K.~T., \& {Goad}, M.~R. 2004, \apj, 606, 749

\bibitem[{{Korista} {et~al.}(1995){Korista}, {Alloin}, {Barr}, {Clavel},
  {Cohen}, {Crenshaw}, {Evans}, {Horne}, {Koratkar}, {Kriss}, {Krolik},
  {Malkan}, {Morris}, {Netzer}, {O'Brien}, {Peterson}, {Reichert},
  {Rodriguez-Pascual}, {Wamsteker}, {Anderson}, {Axon}, {Benitez}, {Berlind},
  {Bertram}, {Blackwell}, {Bochkarev}, {Boisson}, {Carini}, {Carrillo},
  {Carone}, {Cheng}, {Christensen}, {Chuvaev}, {Dietrich}, {Dokter},
  {Doroshenko}, {Dultzin-Hacyan}, {England}, {Espey}, {Filippenko}, {Gaskell},
  {Goad}, {Ho}, {Huchra}, {Jiang}, {Kaspi}, {Kollatschny}, {Laor}, {Luminet},
  {MacAlpine}, {MacKenty}, {Malkov}, {Maoz}, {Martin}, {Matheson}, {McCollum},
  {Merkulova}, {Metik}, {Mignoli}, {Miller}, {Pastoriza}, {Pelat}, {Penfold},
  {Perez}, {Perola}, {Persaud}, {Peters}, {Pitts}, {Pogge}, {Pronik}, {Pronik},
  {Ptak}, {Rawley}, {Recondo-Gonzalez}, {Rodriguez-Espinosa}, {Romanishin},
  {Sadun}, {Salamanca}, {Santos-Lleo}, {Sekiguchi}, {Sergeev}, {Shapovalova},
  {Shields}, {Shrader}, {Shull}, {Silbermann}, {Sitko}, {Skillman}, {Smith},
  {Smith}, {Snijders}, {Sparke}, {Stirpe}, {Stoner}, {Sun}, {Thiele}, {Tokarz},
  {Tsvetanov}, {Turnshek}, {Veilleux}, {Wagner}, {Wagner}, {Wanders}, {Wang},
  {Welsh}, {Weymann}, {White}, {Wilkes}, {Wills}, {Winge}, {Wu}, \&
  {Zou}}]{korista95}
{Korista}, K.~T., {Alloin}, D., {Barr}, P., {et~al.} 1995, \apjs, 97, 285

\bibitem[{{Koz{\l}owski} {et~al.}(2010){Koz{\l}owski}, {Kochanek}, {Udalski},
  {Wyrzykowski}, {Soszy{\'n}ski}, {Szyma{\'n}ski}, {Kubiak}, {Pietrzy{\'n}ski},
  {Szewczyk}, {Ulaczyk}, {Poleski}, \& {OGLE Collaboration}}]{kozlowski10}
{Koz{\l}owski}, S., {Kochanek}, C.~S., {Udalski}, A., {et~al.} 2010, \apj, 708,
  927

\bibitem[{{Krolik} \& {Done}(1995)}]{krolik95}
{Krolik}, J.~H., \& {Done}, C. 1995, \apj, 440, 166

\bibitem[{{Li} {et~al.}(2013){Li}, {Wang}, {Ho}, {Du}, \& {Bai}}]{li13}
{Li}, Y.-R., {Wang}, J.-M., {Ho}, L.~C., {Du}, P., \& {Bai}, J.-M. 2013, \apj,
  779, 110

\bibitem[{{Lynden-Bell} \& {Rees}(1971)}]{lynden71}
{Lynden-Bell}, D., \& {Rees}, M.~J. 1971, \mnras, 152, 461

\bibitem[{{MacLeod} {et~al.}(2010){MacLeod}, {Ivezi{\'c}}, {Kochanek},
  {Koz{\l}owski}, {Kelly}, {Bullock}, {Kimball}, {Sesar}, {Westman}, {Brooks},
  {Gibson}, {Becker}, \& {de Vries}}]{macleod10}
{MacLeod}, C.~L., {Ivezi{\'c}}, {\v Z}., {Kochanek}, C.~S., {et~al.} 2010,
  \apj, 721, 1014

\bibitem[{{Marconi} {et~al.}(2008){Marconi}, {Axon}, {Maiolino}, {Nagao},
  {Pastorini}, {Pietrini}, {Robinson}, \& {Torricelli}}]{marconi08}
{Marconi}, A., {Axon}, D.~J., {Maiolino}, R., {et~al.} 2008, \apj, 678, 693

\bibitem[{{Marconi} {et~al.}(2009){Marconi}, {Axon}, {Maiolino}, {Nagao},
  {Pietrini}, {Risaliti}, {Robinson}, \& {Torricelli}}]{marconi09}
---. 2009, \apjl, 698, L103

\bibitem[{{Morgan} {et~al.}(2010){Morgan}, {Kochanek}, {Morgan}, \&
  {Falco}}]{morgan10}
{Morgan}, C.~W., {Kochanek}, C.~S., {Morgan}, N.~D., \& {Falco}, E.~E. 2010,
  \apj, 712, 1129

\bibitem[{{Mushotzky} {et~al.}(2011){Mushotzky}, {Edelson}, {Baumgartner}, \&
  {Gandhi}}]{mushotzky11}
{Mushotzky}, R.~F., {Edelson}, R., {Baumgartner}, W., \& {Gandhi}, P. 2011,
  \apjl, 743, L12

\bibitem[{{Netzer}(2009)}]{netzer09}
{Netzer}, H. 2009, \apj, 695, 793

\bibitem[{{Netzer} \& {Marziani}(2010)}]{netzer10}
{Netzer}, H., \& {Marziani}, P. 2010, \apj, 724, 318

\bibitem[{{Onken} {et~al.}(2004){Onken}, {Ferrarese}, {Merritt}, {Peterson},
  {Pogge}, {Vestergaard}, \& {Wandel}}]{onken04}
{Onken}, C.~A., {Ferrarese}, L., {Merritt}, D., {et~al.} 2004, \apj, 615, 645

\bibitem[{{Pancoast} {et~al.}(2011){Pancoast}, {Brewer}, \&
  {Treu}}]{pancoast11}
{Pancoast}, A., {Brewer}, B.~J., \& {Treu}, T. 2011, \apj, 730, 139

\bibitem[{{Pancoast} {et~al.}(2014){Pancoast}, {Brewer}, {Treu}, {Park},
  {Barth}, {Bentz}, \& {Woo}}]{pancoast14}
{Pancoast}, A., {Brewer}, B.~J., {Treu}, T., {et~al.} 2014, \mnras, 445, 3073

\bibitem[{{Pancoast} {et~al.}(2012){Pancoast}, {Brewer}, {Treu}, {Barth},
  {Bennert}, {Canalizo}, {Filippenko}, {Gates}, {Greene}, {Li}, {Malkan},
  {Sand}, {Stern}, {Woo}, {Assef}, {Bae}, {Buehler}, {Cenko}, {Clubb},
  {Cooper}, {Diamond-Stanic}, {Hiner}, {H{\"o}nig}, {Joner}, {Kandrashoff},
  {Laney}, {Lazarova}, {Nierenberg}, {Park}, {Silverman}, {Son}, {Sonnenfeld},
  {Thorman}, {Tollerud}, {Walsh}, \& {Walters}}]{pancoast12}
---. 2012, \apj, 754, 49

\bibitem[{{Park} {et~al.}(2012{\natexlab{a}}){Park}, {Kelly}, {Woo}, \&
  {Treu}}]{park12b}
{Park}, D., {Kelly}, B.~C., {Woo}, J.-H., \& {Treu}, T. 2012{\natexlab{a}},
  \apjs, 203, 6

\bibitem[{{Park} {et~al.}(2012{\natexlab{b}}){Park}, {Woo}, {Treu}, {Barth},
  {Bentz}, {Bennert}, {Canalizo}, {Filippenko}, {Gates}, {Greene}, {Malkan}, \&
  {Walsh}}]{park12a}
{Park}, D., {Woo}, J.-H., {Treu}, T., {et~al.} 2012{\natexlab{b}}, \apj, 747,
  30

\bibitem[{{Peterson}(1993)}]{peterson93}
{Peterson}, B.~M. 1993, \pasp, 105, 247

\bibitem[{{Peterson} {et~al.}(1998){Peterson}, {Wanders}, {Horne}, {Collier},
  {Alexander}, {Kaspi}, \& {Maoz}}]{peterson98}
{Peterson}, B.~M., {Wanders}, I., {Horne}, K., {et~al.} 1998, \pasp, 110, 660

\bibitem[{{Peterson} {et~al.}(1991){Peterson}, {Balonek}, {Barker}, {Bechtold},
  {Bertram}, {Bochkarev}, {Bolte}, {Bond}, {Boroson}, {Carini}, {Carone},
  {Christensen}, {Clements}, {Cochran}, {Cohen}, {Crampton}, {Dietrich},
  {Elvis}, {Ferguson}, {Filippenko}, {Fricke}, {Gaskell}, {Halpern}, {Huchra},
  {Hutchings}, {Kollatschny}, {Koratkar}, {Korista}, {Krolik}, {Lame}, {Laor},
  {Leacock}, {MacAlpine}, {Malkan}, {Maoz}, {Miller}, {Morris}, {Netzer},
  {Oliveira}, {Penfold}, {Penston}, {Perez}, {Pogge}, {Richmond}, {Romanishin},
  {Rosenblatt}, {Saddlemyer}, {Sadun}, {Sawyer}, {Shields}, {Shapovalova},
  {Smith}, {Smith}, {Smith}, {Sun}, {Thiele}, {Turner}, {Veilleux}, {Wagner},
  {Weymann}, {Wilkes}, {Wills}, {Wills}, \& {Younger}}]{peterson91}
{Peterson}, B.~M., {Balonek}, T.~J., {Barker}, E.~S., {et~al.} 1991, \apj, 368,
  119

\bibitem[{{Peterson} {et~al.}(2004){Peterson}, {Ferrarese}, {Gilbert}, {Kaspi},
  {Malkan}, {Maoz}, {Merritt}, {Netzer}, {Onken}, {Pogge}, {Vestergaard}, \&
  {Wandel}}]{peterson04}
{Peterson}, B.~M., {Ferrarese}, L., {Gilbert}, K.~M., {et~al.} 2004, \apj, 613,
  682

\bibitem[{{Peterson} {et~al.}(2013){Peterson}, {Denney}, {De Rosa}, {Grier},
  {Pogge}, {Bentz}, {Kochanek}, {Vestergaard}, {Kilerci-Eser}, {Dalla
  Bont{\`a}}, \& {Ciroi}}]{peterson13}
{Peterson}, B.~M., {Denney}, K.~D., {De Rosa}, G., {et~al.} 2013, \apj, 779,
  109

\bibitem[{{Proga}(1999)}]{proga99}
{Proga}, D. 1999, \mnras, 304, 938

\bibitem[{{Urry} \& {Padovani}(1995)}]{urry95}
{Urry}, C.~M., \& {Padovani}, P. 1995, \pasp, 107, 803

\bibitem[{{Walsh} {et~al.}(2009){Walsh}, {Minezaki}, {Bentz}, {Barth},
  {Baliber}, {Li}, {Stern}, {Bennert}, {Brown}, {Canalizo}, {Filippenko},
  {Gates}, {Greene}, {Malkan}, {Sakata}, {Street}, {Treu}, {Woo}, \&
  {Yoshii}}]{walsh09}
{Walsh}, J.~L., {Minezaki}, T., {Bentz}, M.~C., {et~al.} 2009, \apjs, 185, 156

\bibitem[{{Wandel} {et~al.}(1999){Wandel}, {Peterson}, \& {Malkan}}]{wandel99}
{Wandel}, A., {Peterson}, B.~M., \& {Malkan}, M.~A. 1999, \apj, 526, 579

\bibitem[{{Welsh}(1999)}]{welsh99}
{Welsh}, W.~F. 1999, \pasp, 111, 1347

\bibitem[{{Whittle}(1992)}]{whittle92}
{Whittle}, M. 1992, \apjs, 79, 49

\bibitem[{{Woo} {et~al.}(2013){Woo}, {Schulze}, {Park}, {Kang}, {Kim}, \&
  {Riechers}}]{woo13}
{Woo}, J.-H., {Schulze}, A., {Park}, D., {et~al.} 2013, \apj, 772, 49

\bibitem[{{Woo} {et~al.}(2010){Woo}, {Treu}, {Barth}, {Wright}, {Walsh},
  {Bentz}, {Martini}, {Bennert}, {Canalizo}, {Filippenko}, {Gates}, {Greene},
  {Li}, {Malkan}, {Stern}, \& {Minezaki}}]{woo10}
{Woo}, J.-H., {Treu}, T., {Barth}, A.~J., {et~al.} 2010, \apj, 716, 269

\bibitem[{{Zu} {et~al.}(2013){Zu}, {Kochanek}, {Koz{\l}owski}, \&
  {Udalski}}]{zu13}
{Zu}, Y., {Kochanek}, C.~S., {Koz{\l}owski}, S., \& {Udalski}, A. 2013, \apj,
  765, 106

\bibitem[{{Zu} {et~al.}(2011){Zu}, {Kochanek}, \& {Peterson}}]{zu11}
{Zu}, Y., {Kochanek}, C.~S., \& {Peterson}, B.~M. 2011, \apj, 735, 80

\end{thebibliography}

\end{document}